\documentclass[aps,prb,reprint,floatfix]{revtex4-2}

\usepackage{iftex}
\ifPDFTeX
  \ifnum\pdfoutput>0
    \usepackage{graphicx}
    \usepackage{xcolor}
  \else
    \usepackage[dvipdfmx]{graphicx}
    \usepackage[dvipdfmx]{xcolor}
  \fi
\else
  \usepackage{graphicx}
  \usepackage{xcolor}
\fi
\usepackage{amsmath,amssymb,amsfonts}
\usepackage{mathrsfs}
\usepackage{mathtools}
\usepackage{bm}
\usepackage{placeins}

\def\e{{\epsilon}}
\def\k{{{{\bm k}}}}

\def\q{{{{\bm q}}}}

\def\a{{\alpha}}
\def\b{{\beta}}

\allowdisplaybreaks[4]
\begin{document}

\title{
Quantum-interference-driven orbital density wave and high-temperature superconductivity in trilayer nickelates
}

\author{Youichi Yamakawa}
\email{yamakawa.youichi.p1@f.mail.nagoya-u.ac.jp}
\author{Hiroshi Kontani}
\email{kontani.hiroshi.v4@f.mail.nagoya-u.ac.jp}
\affiliation{Department of Physics, Nagoya University, Furo-cho, Nagoya 464-8602, Japan}
\date{\today}

\begin{abstract}
Intertwined charge-density-wave (CDW) and spin-density-wave (SDW) orders are a hallmark of high-temperature superconducting multilayer nickelates.
In trilayer La$_4$Ni$_3$O$_{10}$, charge correlations develop at temperatures above the onset of long-range spin order, and the characteristic ordering wavevectors satisfy ${\bm Q}_{\rm cdw}\approx2{\bm Q}_{\rm sdw}$.
Here, using a density-wave equation with vertex corrections, we show that quantum interference between short-range SDW fluctuations at ${\bm q}\approx{\bm Q}_{\rm sdw}$ on the outer NiO$_2$ layers generates an inter-outer-layer bond order at
${\bm Q}_{\rm cdw}\approx2{\bm Q}_{\rm sdw}$.
This bond order induces a pronounced inner-layer-centered orbital order, with antiphase modulations of the Ni $d_{3z^2-r^2}$ and $d_{x^2-y^2}$ occupations, producing strong orbital polarization but only weak total charge modulation.
This intertwined bond-and-orbital order accounts for the layer-selective electronic reconstruction inferred from NMR/NQR and is consistent with Raman spectroscopy and scanning tunnelling microscopy measurements.
The same orbital and spin fluctuations also cooperate to stabilize $s_{\pm}$-wave superconductivity through $M_z$ mirror-parity selection rules.
Our results provide a unified microscopic framework for intertwined density-wave order and high-$T_c$ superconductivity in multilayer nickelates.

\end{abstract}

\maketitle

\section{Introduction}

The discovery of high-temperature superconductivity in multilayer Ruddlesden--Popper nickelates has opened a new avenue for unconventional superconductivity beyond the cuprates. In bilayer La$_3$Ni$_2$O$_7$, superconductivity approaching 80~K has been reported under high pressure~\cite{E1-Sun}, while trilayer La$_4$Ni$_3$O$_{10}$ also becomes superconducting under pressure at
temperatures of several tens of kelvin~\cite{E2-Zhu,E3-Zhang}.
Their low-energy electronic structures are dominated by the Ni $d_{x^2-y^2}$ and $d_{3z^2-r^2}$ orbitals and are shaped by $d_{3z^2-r^2}$ orbital interlayer hybridization~\cite{E10-Li,T11-Chen}.
Understanding how these ingredients give rise to high-$T_c$ superconductivity is therefore a central issue in correlated-electron physics.

A key clue lies in the density-wave states that emerge near superconductivity.
Both charge-density-wave (CDW) and spin-density-wave (SDW) orders have been observed in multilayer nickelates~\cite{E4-Zhang,E5-Wang,E7-Li},
and superconductivity emerges as these orders are suppressed under pressure~\cite{E2-Zhu,E3-Zhang,E9-Xu}.
Theoretical studies have explored spin-fluctuation mechanisms \cite{T1-Sakakibara,T3-Zhang,T4-Zhang,T5-Zhang},
functional renormalization-group approaches~\cite{T2-Yang},
interlayer pairing and strong-coupling models~\cite{T8-Huang,T9-Oh},
and orbital-selective correlations~\cite{T6-Leonov,T7-Wang}.
\textcolor{black}{Spin--charge--orbital ordered states have also been investigated using symmetry analysis and density functional theory~\cite{T10-Zhang}.}
Yet the microscopic origin of the density-wave orders, particularly the CDW, remains unresolved.
In conventional mean-field and random-phase-approximation (RPA) treatments with local Coulomb interactions, Fermi-surface nesting readily promotes an SDW instability, whereas a comparably strong CDW instability is difficult to obtain.

The close connection between CDW and SDW orders in multilayer nickelates is reminiscent of the successive electronic nematic (${\bm q}_{\rm nem}={\bm 0}$) and SDW transitions in Fe-based superconductors~\textcolor{black}{\cite{T16-Fernandes,T17-Yamakawa}}.
In nickelates, however, the charge order develops at a finite wavevector, ${\bm Q}_{\rm cdw}\neq{\bm 0}$, making its microscopic origin even more nontrivial.
A theoretical framework for addressing this problem is provided by the density-wave (DW) equation, which incorporates vertex corrections describing quantum interference between paramagnons, collective spin fluctuations in the absence of long-range magnetic order~\cite{T15-Kontani,T24-Tsuchiizu,T12-Tazai,T18-Tazai}.
Quantum interference is particularly pronounced in low-dimensional systems.

In this framework, interference between spin fluctuations near $\pm{\bm Q}_{\rm sdw}$ can drive charge-channel instabilities at ${\bm q}\sim2{\bm Q}_{\rm sdw}$ and ${\bm q}\sim{\bm 0}$.
Trilayer La$_4$Ni$_3$O$_{10}$ provides an ideal setting to test this mechanism.
Scattering experiments have established the characteristic relation~\cite{E4-Zhang}
\begin{equation}
{\bm Q}_{\rm cdw}\approx2{\bm Q}_{\rm sdw},
\label{eq:Qrelation-exp}
\end{equation}
while site-selective $^{139}$La NMR/NQR measurements show that a charge-related anomaly develops predominantly in the inner NiO$_2$ layer below $T^*\sim150$~K, preceding long-range SDW order at $T_{\rm SDW}\sim133$~K~\cite{E5-Wang}.
Polarized Raman spectroscopy reveals a pronounced multiorbital character of the density-wave reconstruction~\cite{E6-Suthar}, while scanning tunnelling microscopy directly visualizes the
incommensurate unidirectional CDW in real space~\cite{E7-Li}.
These observations call for a microscopic mechanism that accounts for both the CDW--SDW wavevector relation and the layer- and orbital-dependent electronic reconstruction.

Here, applying the DW equation to a realistic trilayer multiorbital model, we show that quantum interference between spin fluctuations on the outer NiO$_2$ layers (OLs) generates an inter-OL bond order at ${\bm Q}_{\rm cdw}\approx2{\bm Q}_{\rm sdw}$.
Unexpectedly, this bond order in turn induces a pronounced orbital order in the inner NiO$_2$ layer (IL), characterized by antiphase modulations of the Ni $d_{3z^2-r^2}$ and $d_{x^2-y^2}$ occupations.
As a result, a large orbital polarization develops with only weak net on-site charge modulation.
This intertwined bond-and-orbital order accounts for the layer-selective electronic reconstruction inferred from NMR/NQR and is consistent with Raman and scanning tunnelling microscopy.
Furthermore, the CDW form factor is even under the mirror operation $M_z$ about the IL, whereas the SDW form factor is odd.
The resulting selection rules allow CDW and SDW fluctuations to cooperate in stabilizing $s_{\pm}$-wave superconductivity.
Our results provide a unified microscopic framework for intertwined density-wave order and high-$T_c$ superconductivity in multilayer nickelates.

\section{Results and Discussion}

\subsection{Multilayer model and electronic structure}

\begin{figure}[!tbp]
\centering
\includegraphics[width=\linewidth]{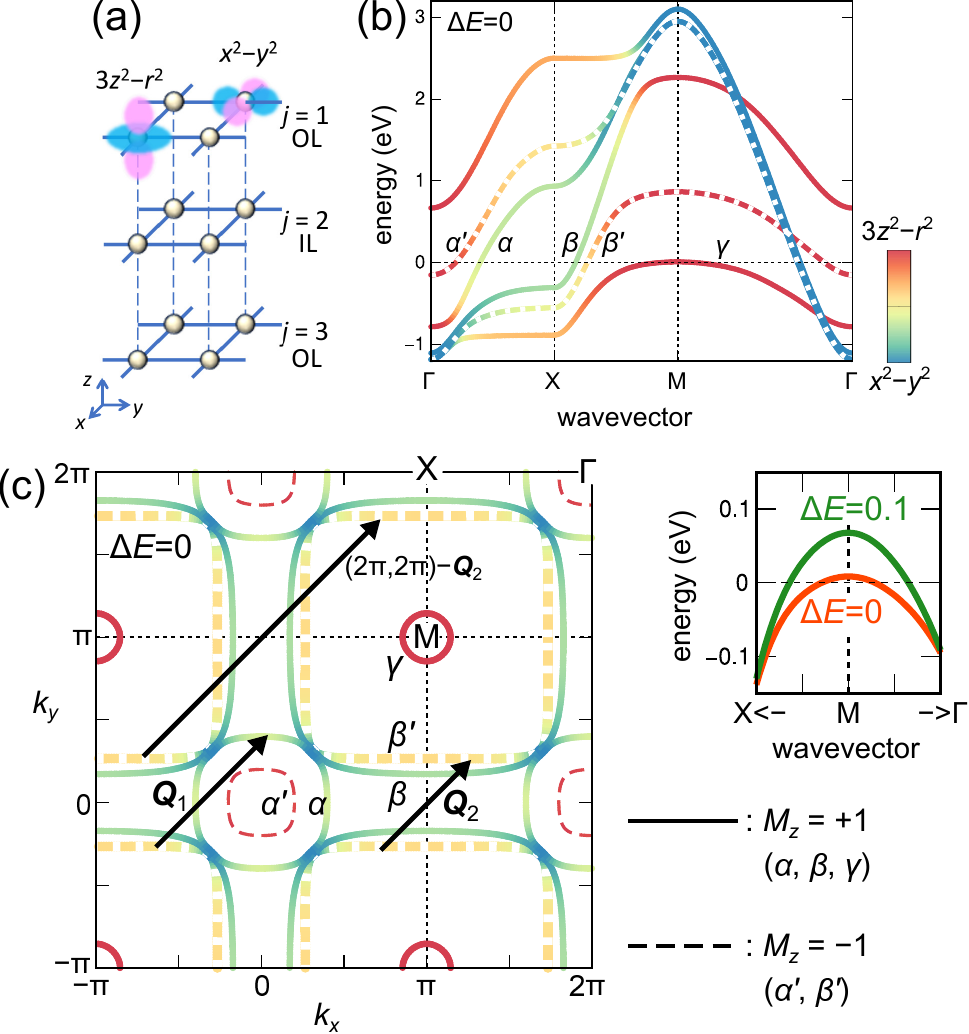}
\caption{
{\bf Electronic structure of trilayer La$_4$Ni$_3$O$_{10}$:} \ 
(a) Trilayer crystal structure and the two Ni $e_g$ orbitals, $d_{3z^2-r^2}$ and $d_{x^2-y^2}$.
The three NiO$_2$ layers are labelled $j=1,2,3$, corresponding to the OL--IL--OL structure.
(b) First-principles-based band structure of the trilayer two-orbital model.
The $d_{3z^2-r^2}$ ($d_{x^2-y^2}$) orbital weight is expressed by red (green) color. 
Solid and dashed lines denote the $M_z=+1$ and $-1$ sectors, respectively. 
(c) Fermi surfaces with the $d_{3z^2-r^2}$ and $d_{x^2-y^2}$ orbital weights.
${\bm Q}_1$ and ${\bm Q}_2$ denote the characteristic nesting wavevectors.
Inset: Dependence of the $\gamma$ band near the M point on $\Delta E$.
}
\label{fig1}
\end{figure}

Figure~\ref{fig1}(a) shows the trilayer crystal structure of La$_4$Ni$_3$O$_{10}$ and the two Ni $e_g$ orbitals, $d_{3z^2-r^2}$ and $d_{x^2-y^2}$, that dominate the low-energy electronic states.
We label the three NiO$_2$ layers by $j=1,2,3$, with $j=1,3$ corresponding to the OLs and $j=2$ to the IL.
Unless otherwise stated, energies and temperatures are expressed in eV, with $k_{\rm B}=1$.

We construct a first-principles-based six-band model comprising the two Ni $e_g$ orbitals on each of the three layers. 
The model is based on the experimentally determined tetragonal $I4/mmm$ crystal structure of La$_4$Ni$_3$O$_{10}$ at $P=16.6$~GPa~\cite{E11-Li}; see Sec.~\ref{sec:methodA}.
The resulting band structure and Fermi surfaces are shown in Figs.~\ref{fig1}(b) and \ref{fig1}(c).
The Fermi surfaces consist of the electron pockets $\alpha$ and $\alpha'$, and the hole pockets $\beta$, $\beta'$, and $\gamma$.
ARPES measurements also indicate an $\alpha'$ electron band near the Fermi level~\cite{E10-Li}.
The characteristic nesting vectors ${\bm Q}_1$ and ${\bm Q}_2$ are indicated in Fig.~\ref{fig1}(c).

The trilayer structure is invariant under the mirror operation $M_z:z\rightarrow-z$ about the IL.
The electronic states can therefore be classified by their mirror eigenvalues $\pm1$.
The $\alpha$, $\beta$, and $\gamma$ pockets belong to the mirror-even ($M_z=+1$) sector, whereas the $\alpha'$ and $\beta'$ pockets belong to the mirror-odd ($M_z=-1$) sector.
The nesting vector ${\bm Q}_1$ connects Fermi surfaces with opposite mirror parities, whereas ${\bm Q}_2$ connects those with the same mirror parity. As shown below, ${\bm Q}_{\rm sdw}\approx{\bm Q}_1$ and ${\bm Q}_{\rm cdw}\approx{\bm Q}_2$.
This parity structure will play a key role in both the density-wave instabilities and superconducting pairing.

The first-principles model contains a very small $\gamma$ pocket around the M point.
For the CDW-instability analysis, the normal-state electronic structure above the density-wave transition is particularly relevant.
Temperature-dependent ARPES measurements show that the $d_{3z^2-r^2}$-dominated $\gamma$ band crosses the Fermi level above the density-wave transition~\cite{E10-Li}, whereas more recent measurements find a nearly temperature-independent flat $\gamma$ band located very close to the Fermi level \cite{E14-Du}.
To examine the sensitivity of our results to the position of the $\gamma$ band, we introduce an energy shift $\Delta E$ around the M point.

We treat electronic correlations within the multiorbital fluctuation-exchange (FLEX) approximation and use the resulting dressed Green functions consistently in the subsequent analyses.
We investigate the CDW instability using a conserving DW equation with vertex corrections and superconducting pairing using the FLEX and Bethe--Salpeter (BS) approaches.
Details are given in Secs.~\ref{sec:methodB}--\ref{sec:methodD}.

\subsection{SDW fluctuations and their odd-parity form factor}

\begin{figure*}[!tp]
\centering
\includegraphics[width=0.70\textwidth]{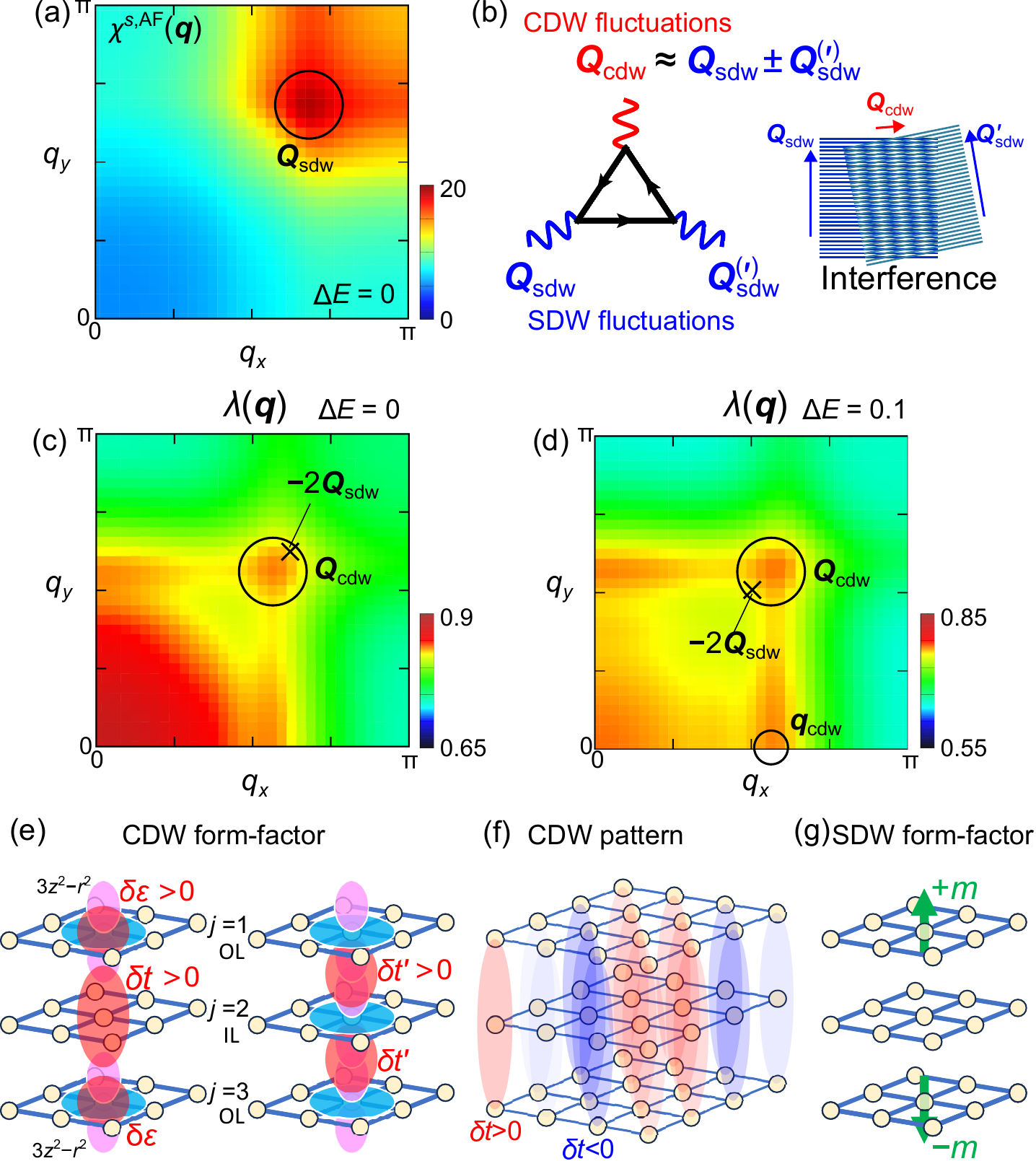}
\caption{
{\bf Spin fluctuations and quantum-interference-driven density-wave instability:} \ 
(a) Inter-OL antiphase spin susceptibility $\chi^{s,{\rm AF}}({\bm q})$ obtained by FLEX.
(b) Representative quantum-interference process in which two paramagnons generate charge and orbital fluctuations. The solid lines represent electronic Green functions.
(c) Obtained $\lambda({\bm q})$ of the charge-channel DW equation for $\Delta E=0$.
The finite-wavevector instability appears at ${\bm Q}_{\rm cdw}\sim{\bm Q}_2$ and satisfies ${\bm Q}_{\rm cdw}\approx-2{\bm Q}_{\rm sdw}$ modulo a reciprocal-lattice vector.
(d) Corresponding result for $\Delta E=0.1$, where the $\gamma$ pocket is enlarged.
(e) Dominant real-space components of the CDW form factor, including the inter-OL bond modulation $\delta t$, the IL-OL bond modulation $\delta t'$, and the OL local-potential modulation $\delta\varepsilon$.
(f) Real-space structure of the inter-OL bond order at ${\bm Q}_{\rm cdw}$.
(g) Layer-dependent SDW form factor with antiphase moments on the two OLs.
In {\bf a,c,d}, we set $U=3$ and $T=0.01$.
}
\label{fig2}
\end{figure*}

Within FLEX, we calculate the full spin-susceptibility tensor $\chi^s_{mm';ll'}(q)$, where $m=(d,j)$ is a composite orbital--layer index.
Our numerical calculations show that the leading spin-fluctuation mode is odd under the exchange of the two OLs.
To characterize these antiferromagnetically (AF) correlated spin fluctuations, we project the susceptibility onto the layer form factor
\begin{equation}
f^{\rm AF}(j)=2-j ,
\end{equation}
which corresponds to $(1,0,-1)$ for $j=(1,2,3)$.
This form factor is odd under the layer exchange $j\rightarrow4-j$ and describes spin fluctuations with opposite signs on the two OLs and zero amplitude on the IL.
The spin susceptibility in this odd-parity channel is defined as
\begin{equation}
\chi^{s,{\rm AF}}(q)
=
\sum_{m,l}
\chi^s_{m,m;l,l}(q)
f^{\rm AF}(j_m)f^{\rm AF}(j_l),
\label{eq:chiAF}
\end{equation}
where $j_m$ denotes the layer component of the composite
index $m$.

Figure~\ref{fig2}(a) shows the static susceptibility $\chi^{s,{\rm AF}}({\bm q})$.
A pronounced peak appears at ${\bm q}={\bm Q}_{\rm sdw}$, close to the nesting vector ${\bm Q}_1$ in Fig.~\ref{fig1}(c).
An equally strong peak appears at the symmetry-related wavevector ${\bm Q}'_{\rm sdw}$, obtained by a $90^\circ$ rotation of ${\bm Q}_{\rm sdw}$ and close to ${\bm Q}'_1$.
The obtained spin Stoner factor is $\alpha_S\approx0.945$. 
The enhancement of the leading spin mode is governed by $(1-\alpha_S)^{-1}$, with $\alpha_S=1$ marking the SDW instability.

At $\q={\bm Q}_{\rm sdw}$, the orbital-diagonal contribution from $d_{3z^2-r^2}$ is approximately three times that from $d_{x^2-y^2}$.
This ratio increases as $\alpha_S$ approaches unity in our calculations, indicating increasingly $d_{3z^2-r^2}$-dominated spin fluctuations.
This layer-selective magnetic structure is consistent with previous FRG and RPA analyses of the trilayer system \cite{T2-Yang,T3-Zhang,T5-Zhang}.

\subsection{CDW instability induced by quantum interference mechanism}

We next investigate the CDW instability generated by the quantum-interference mechanism.
Within FLEX alone, the local Coulomb interactions strongly enhance the SDW fluctuations but do not produce a comparably strong CDW instability.
Including the Aslamazov--Larkin (AL) vertex corrections describing quantum interference between paramagnons strongly enhances the charge channel
\cite{T21-Onari,T15-Kontani,T12-Tazai,T19-Onari,T13-Inoue,T22-Yamakawa}.

A representative quantum-interference process is illustrated in Fig.~\ref{fig2}(b).
Interference between paramagnons near $\pm{\bm Q}_{\rm sdw}$ and $\pm{\bm Q}'_{\rm sdw}$ enhances charge-channel fluctuations at
\begin{equation}
{\bm q}
\sim
{\bm 0},\quad
\pm2{\bm Q}_{\rm sdw},\quad 
\pm({\bm Q}_{\rm sdw}+{\bm Q}'_{\rm sdw}),
\label{eq:QIwave}
\end{equation}
together with their symmetry-related wavevectors.
We incorporate these processes into the vertex kernel of the multiorbital DW equation and determine its leading eigenvalue and corresponding form factor~\cite{T15-Kontani}.
Details are given in Sec.~\ref{sec:methodC} and Appendix~\ref{app:F}.

Figures~\ref{fig2}(c) and \ref{fig2}(d) show the momentum dependence of the leading charge-channel eigenvalue $\lambda({\bm q})$ for $\Delta E=0$ and $0.1$, respectively.
Peaks appear at the wavevectors expected from Eq.~(\ref{eq:QIwave}).
In particular, a pronounced peak occurs at ${\bm Q}_{\rm cdw}\approx{\bm Q}_2$, satisfying
${\bm Q}_{\rm cdw} \approx \pm2{\bm Q}_{\rm sdw}+{\bm G}$,
where ${\bm G}$ is a reciprocal-lattice vector.
This result provides a microscopic explanation for the enhancement of the charge channel at the experimentally observed ordering wavevector.

The ${\bm Q}_{\rm cdw}$ channel competes closely with those at ${\bm q}={\bm 0}$ and ${\bm q}_{\rm cdw}\equiv {\bm Q}_{\rm sdw}+{\bm Q}'_{\rm sdw}$.
For $\Delta E=0$, $\lambda({\bm 0})\gtrsim\lambda({\bm Q}_{\rm cdw}), \lambda({\bm q}_{\rm cdw})$,
whereas the three peak values are nearly degenerate for $\Delta E=0.1$.
Their relative magnitudes depend on details of the model, including the size of the $\gamma$ pocket, suggesting competing charge-channel ordering tendencies at these wavevectors.
Nevertheless, the peak at ${\bm Q}_{\rm cdw}$ persists even for $\Delta E=-0.1$, where the $\gamma$ pocket is absent (see Appendix~\ref{app:H}).

\subsection{Even-parity CDW form factor}

We next examine the microscopic structure of the leading charge-channel eigenmode at ${\bm Q}_{\rm cdw}$. 
The quantum-interference mechanism can generate diverse correlation-driven orders, including bond, orbital, and charge- and spin-loop-current orders.
\textcolor{black}{An application to charge-loop-current order in kagome metals is given in Ref.~\cite{T20-Tazai}.}
Such orders can develop with little or no modulation of the total on-site electron density, reducing the local Coulomb-energy cost associated with charge disproportionation.

The microscopic structure of the CDW is encoded in the DW-equation eigenfunction, or form factor, $\hat f_{\bm q}(k)$, at ${\bm q}={\bm Q}_{\rm cdw}$.
Its detailed momentum dependence is presented in Appendix~\ref{app:F}.
We use the compact notation $k=({\bm k},\epsilon_n)$ and $q=({\bm q},\omega_\nu)$, where $\epsilon_n=(2n+1)\pi T$ and $\omega_\nu=2\nu\pi T$ are fermionic and bosonic Matsubara frequencies, respectively.

To characterize its real-space structure, we define
\begin{equation}
F_{{\bm q};d_m d_l}^{j_m j_l}({\bm r})
\equiv
\frac{1}{N}\sum_{\bm k}
f_{\bm q}^{ml}({\bm k}-{\bm q}/2)
e^{i{\bm k}\cdot{\bm r}},
\label{eq:realspace-formfactor}
\end{equation}
where $m=(d_m,j_m)$ and $l=(d_l,j_l)$,
${\bm r}$ is the in-plane relative lattice vector, and $N$ is the number of in-plane unit cells.
Here, $d=1,2$ denote the $d_{3z^2-r^2}$ and $d_{x^2-y^2}$ orbitals, respectively.

We now discuss the real-space structure of the form factor at ${\bm q}={\bm Q}_{\rm cdw}$.
The largest component is the vertical inter-OL bond modulation $\delta t\equiv F_{{\bm q};11}^{13}({\bm 0})$.
Additional sizable components are the vertical OL--IL bond modulation $\delta t'
\equiv F_{{\bm q};11}^{12}({\bm 0})$ and the OL local-potential modulation $\delta\varepsilon\equiv F_{{\bm q};11}^{11}({\bm 0})$.
These components are schematically illustrated in Fig.~\ref{fig2}(e).
In the present numerical study, we obtain $(\delta t,\delta t',\delta\varepsilon)=(0.09, 0.018,0.016)$ at zero energy for $f^{\rm max}=0.1$.
The renormalized CDW mixing gap is $\approx 2f^{\rm max}/Z$, where $Z$ is the mass-enhancement factor.
$Z\sim5$ in the present study; see Appendix~\ref{app:E}.

The CDW form factor is even under the mirror operation $M_z$, which exchanges the two OLs.
In particular, the local-potential modulations on layers 1 and 3 are equal, and the inter-OL bond component is symmetric under this exchange.
Figure~\ref{fig2}(f) illustrates the real-space structure of the inter-OL bond order, whose spatial modulation is proportional to $\cos({\bm Q}_{\rm cdw}\cdot{\bm R}+\phi_0)$, where ${\bm R}$ denotes the in-plane bond-center position and $\phi_0$ is the modulation phase.
For comparison, Fig.~\ref{fig2}(g) shows the odd-parity SDW form factor, with opposite spin modulations on the two OLs.
The local charge and orbital responses induced by the CDW form factor will be analyzed below.

\subsection{Correlated growth of CDW and SDW fluctuations}

\begin{figure}[!tbp]
\centering
\includegraphics[width=\linewidth]{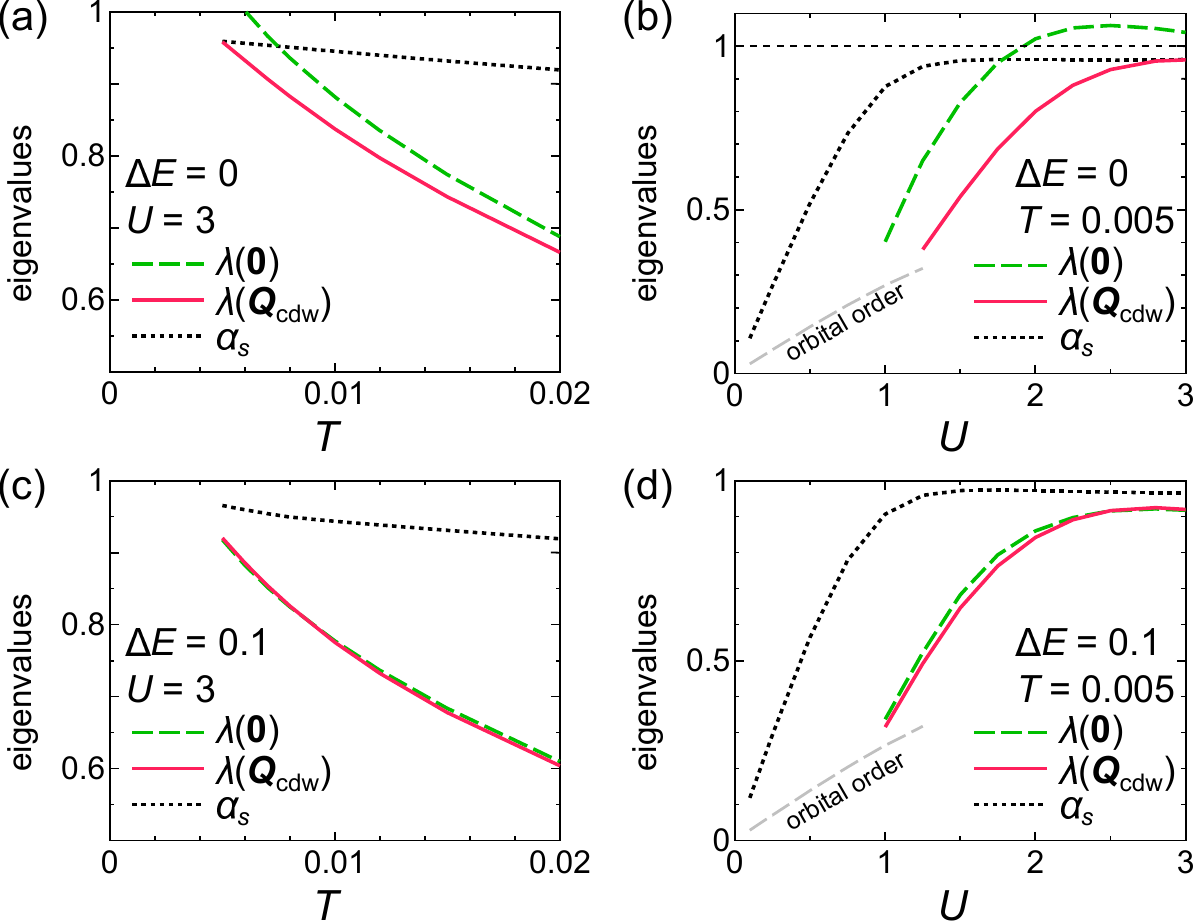}
\caption{
{\bf Development and robustness of the CDW instability:} \ 
(a) Temperature dependences of the leading DW eigenvalues $\lambda({\bm Q}_{\rm cdw})$ and $\lambda({\bm 0})$, and the spin Stoner factor $\alpha_S$, for $\Delta E=0$ and $U=3$.
(b) Corresponding $U$ dependences at $T=0.005$.
(c) and (d) Same as (a) and (b), respectively, but for $\Delta E=0.1$.
}
\label{fig3}
\end{figure}

Figure~\ref{fig3}(a) shows the temperature dependences of the leading DW eigenvalues $\lambda({\bm Q}_{\rm cdw})$ and $\lambda({\bm 0})$, together with the spin Stoner factor $\alpha_S$, for $\Delta E=0$ at fixed $U=3$.
Upon cooling, the increase in $\alpha_S$ is accompanied by a rapid enhancement of $\lambda({\bm Q}_{\rm cdw})$.
A similar trend is found upon increasing $U$ at fixed $T=0.005$, as shown in Fig.~\ref{fig3}(b).
This correlated growth is consistent with the quantum-interference mechanism, in which stronger paramagnon fluctuations enhance the charge-channel instability.

Figures~\ref{fig3}(c) and \ref{fig3}(d) show the corresponding results for $\Delta E=0.1$, where the $\gamma$ pocket is larger. 
Despite quantitative differences, the CDW eigenvalue again increases rapidly as the spin fluctuations grow.
Thus, the connection between the enhanced spin fluctuations and the CDW instability persists upon enlarging the $\gamma$ pocket.

Extrapolating $\lambda({\bm Q}_{\rm cdw})$ to unity gives an estimated $T_{\rm CDW}\sim0.003$--$0.006$~eV ($\sim35$--$70$~K).
Electron--phonon coupling can further enhance this instability~\cite{T23-Kontani}.
For an interlayer $A_{1g}$ phonon mode coupled to the inter-OL bond modulation, we estimate this effect through an approximate additive correction to the DW eigenvalue,
\begin{equation}
\lambda({\bm q})\rightarrow
\lambda({\bm q})+\lambda_{\rm e\text{-}ph},
\end{equation}
where $\lambda_{\rm e\text{-}ph}$ denotes the effective phonon contribution to the relevant DW channel.
Taking $\lambda_{\rm e\text{-}ph}=0.1$ raises the estimated $T_{\rm CDW}$ from the temperature dependences in Figs.~\ref{fig3}(a) and \ref{fig3}(c) to approximately $0.006$--$0.01$~eV ($\sim70$--$115$~K).
Although DFT calculations indicate weak electron--phonon coupling in trilayer nickelates~\cite{T25-Li}, this estimate suggests that even a modest coupling to the bond-order channel can appreciably enhance $T_{\rm CDW}$.

\subsection{Inner-layer orbital-density-wave state}

\begin{figure*}[!tp]
\centering
\includegraphics[width=0.8\textwidth]{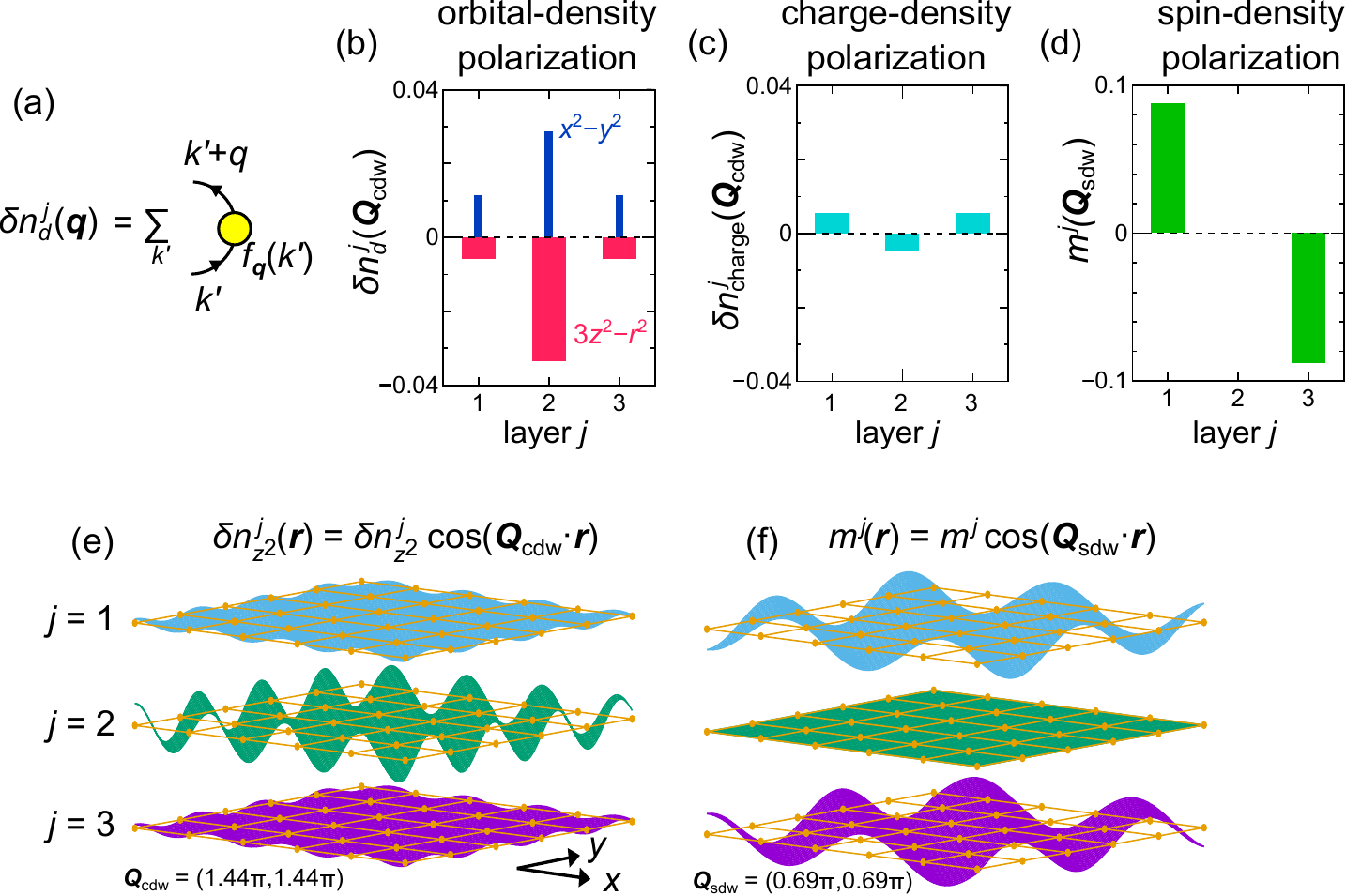}
\caption{
{\bf Orbital-density-wave and spin-density-wave structures:} \ 
(a) Schematic representation of the linear-response process relating the microscopic DW form factor to the induced local density modulation.
(b) Orbital-resolved density modulations induced by the leading CDW form factor. The two Ni $e_g$ occupations modulate predominantly in antiphase, with particularly large amplitudes on the IL.
(c) Corresponding total Ni charge modulation, which remains strongly suppressed.
(d) Layer-resolved SDW amplitude, concentrated on the two OLs with antiphase moments on layers 1 and 3.
(e) and (f) Real-space structures of the orbital density wave at ${\bm Q}_{\rm cdw}$ and the spin density wave at ${\bm Q}_{\rm sdw}$, respectively.
}
\label{fig4}
\end{figure*}

To determine the local charge and orbital responses to the CDW form factor, we calculate the density modulation in each orbital and layer within linear response:
\begin{equation}
\delta n_d^j({\bm q}) =
2\frac{T}{N}\sum_k
\left[
\hat G(k+{\bm q})
\hat f_{\bm q}(k)
\hat G(k)
\right]_{mm},
\label{eq:Dn}
\end{equation}
where $m=(d,j)$.
Here, ${\hat f}_{\bm q}(k)$ is the energy- and momentum-dependent form factor obtained by the DW equation analysis.
This process is illustrated schematically in Fig.~\ref{fig4}(a).
We set ${\bm q}={\bm Q}_{\rm cdw}$ and normalize the form factor to $f^{\rm max}=0.1$ to illustrate the relative orbital and layer amplitudes.

Figures~\ref{fig4}(b) and \ref{fig4}(c) show the orbital-resolved density modulations and the corresponding total Ni charge modulation.
The orbital-resolved modulations are much larger on the IL than on the OLs.
Moreover, the two $e_g$ occupations modulate approximately in antiphase with nearly equal amplitudes,
\begin{equation}
\delta n^{j}_{3z^2-r^2}
\approx
-\delta n^{j}_{x^2-y^2},
\label{eq:delta-n}
\end{equation}
on both the OLs and the IL.
Defining the charge and orbital components as the sum and difference of these two density modulations, respectively, we obtain
\begin{equation}
|\delta n^j_{\rm orbital}|
\gg
|\delta n^j_{\rm charge}|.
\label{eq:delta-n2}
\end{equation}
The induced local density response is therefore dominated by an IL-centered orbital density wave.
The near cancellation of the two orbital contributions reduces the Hartree--Fock energy cost associated with net on-site charge modulation.
This response follows from the DW eigenmode without imposing a local charge-neutrality constraint.

Notably, the IL-centered orbital polarization is induced by a form factor dominated by OL components, demonstrating that the layer structure of the local density response can differ markedly from that of the underlying order parameter.
We have verified that $\delta n^j_{3z^2-r^2}$ in Fig.~\ref{fig4}(b) is reproduced semi-quantitatively by retaining only $\delta t$, $\delta t'$, and $\delta\varepsilon$. 
Also, $\delta n^j_{x^2-y^2}$ is reproduced by retaining the IL nearest-neighbor bond-order, which is given by $F_{{\bm q};22}^{22}({\bm r})=-0.020$ at ${\bm r}=(1,0)$.
Importantly, the IL orbital order remains robust in the present DW equation analysis even in the absence of the $\gamma$ pocket, as shown in Appendix~\ref{app:H}.

The layer-resolved SDW amplitudes in Fig.~\ref{fig4}(d) follow the previously identified odd-parity profile $f^{\rm AF}(j)$.
Thus, the SDW and orbital density wave exhibit complementary layer selectivity, with the former concentrated on the OLs and the latter on the IL.
Figures~\ref{fig4}(e) and \ref{fig4}(f) illustrate the corresponding real-space orbital- and spin-density-wave patterns.
Below the SDW transition, the spin order can additionally induce charge and orbital modulations at $2{\bm Q}_{\rm sdw}$, reinforcing the connection between the two density-wave orders (see Appendix~\ref{app:D}).

\subsection{Comparison with experiments}

The IL-centered orbital polarization obtained here accounts for the layer-selective electronic reconstruction inferred from $^{139}$La NMR/NQR below $T^*\sim150$~K~\cite{E5-Wang}.
The experimentally inferred onset above $T_{\rm SDW}\sim133$~K is consistent with a charge-channel instability driven by paramagnons without requiring long-range spin order.

The predicted orbital density wave is also consistent with the multiorbital reconstruction inferred from polarized Raman spectroscopy~\cite{E6-Suthar} and the incommensurate unidirectional modulation observed by STM~\cite{E7-Li}.
These observations support distinct aspects of the predicted state, although neither directly establishes the antiphase modulation of the two $e_g$ occupations.

The relation ${\bm Q}_{\rm cdw}\approx2{\bm Q}_{\rm sdw}$ established by x-ray and neutron diffraction \cite{E4-Zhang} is consistent with the wavevector selection discussed above.
However, non-resonant x-ray diffraction alone does not uniquely distinguish charge-dominant from orbital-dominant order.
Polarization-resolved resonant x-ray scattering at the Ni edge could provide a more direct test of the predicted orbital modulation.

Our model uses the tetragonal structure at 16.6~GPa, whereas $T^*\sim150$~K is measured at ambient pressure. 
The lower calculated $T_{\rm CDW}$ is therefore compatible with the pressure-induced suppression of density-wave order.
The calculated superconducting $T_c$ on the 100-K scale indicates strong pairing rather than quantitative agreement with experiment.
A quantitative account of the transition temperatures requires a fuller treatment of pressure dependence, electron--phonon coupling, and magnetic coupling between trilayer units; the latter can produce a finite $T_{\rm SDW}$, which is absent in the present two-dimensional FLEX calculation~\cite{T26-Kontani}.

\subsection{Superconductivity}

\begin{figure*}[!tp]
\centering
\includegraphics[width=0.70\textwidth]{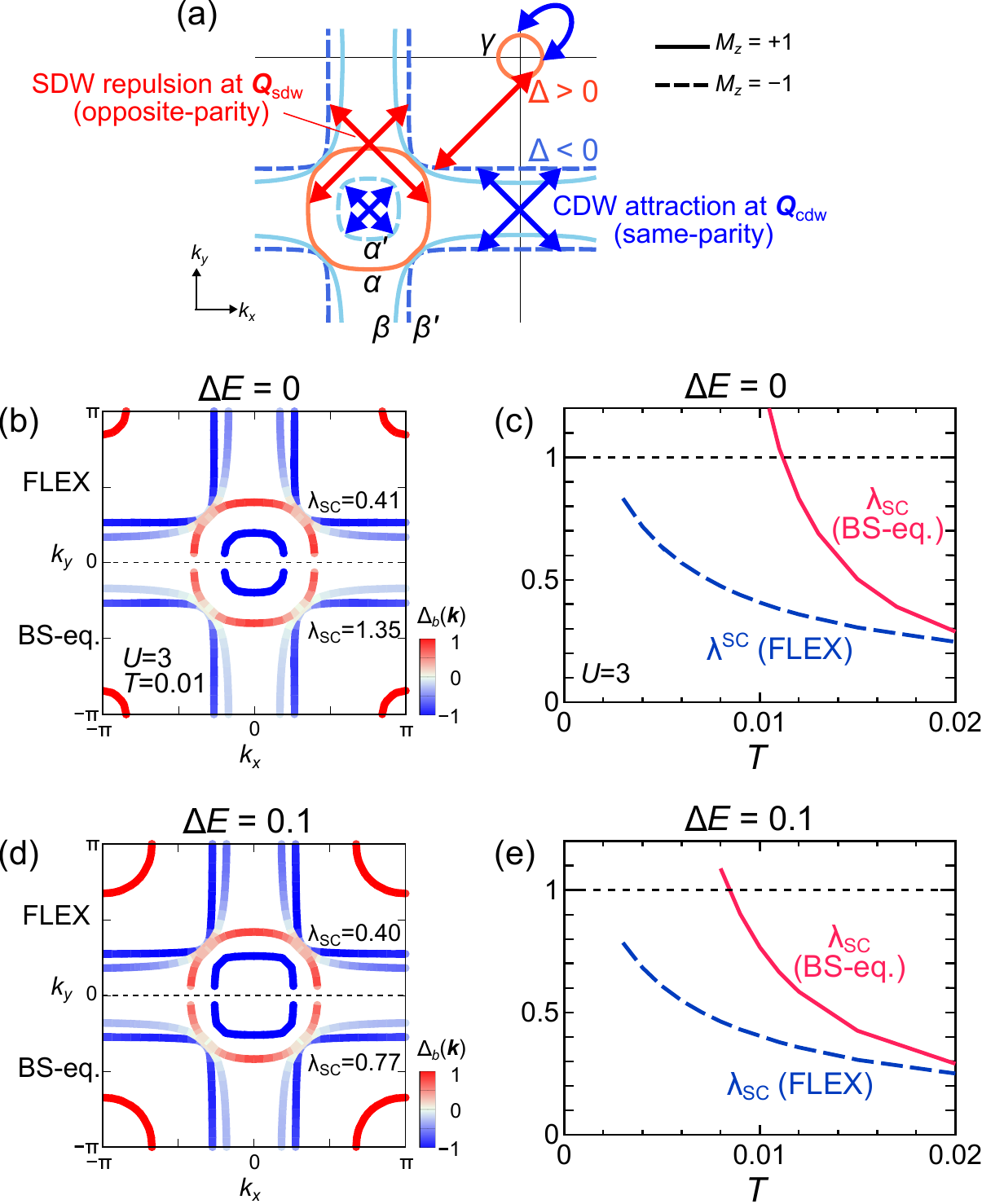}
\caption{
{\bf Cooperative superconducting pairing mediated by CDW and SDW fluctuations:} \ 
(a) Schematic illustration of the $M_z$-parity selection rules for pairing interactions.
CDW fluctuations mainly mediate attraction between Fermi pockets with the same $M_z$ parity, whereas SDW fluctuations mediate repulsion between pockets with opposite parity.
Red and blue indicate opposite signs of the $s_{\pm}$-wave gap.
Solid and dashed lines denote the $M_z=+1$ and $-1$ sectors, respectively. 
(b) Superconducting gap functions obtained from the FLEX and BS-equation analyses for $\Delta E=0$, $U=3$, and $T=0.01$.
(c) Temperature dependence of the corresponding superconducting eigenvalues $\lambda_{\rm SC}$.
(d) and (e) Same as (b) and (c), respectively, but for $\Delta E=0.1$.
These complementary pairing interactions cooperatively stabilize the $s_{\pm}$-wave superconducting state.
}
\label{fig5}
\end{figure*}

Experimentally, pressure suppresses density-wave order and induces superconductivity~\cite{E2-Zhu,E3-Zhang}.
Ultrafast spectroscopy further reports the collapse of the density-wave gap near the pressure range where a superconducting-like low-energy gap emerges \cite{E9-Xu}.
We therefore examine how the CDW and SDW fluctuations identified above contribute to superconducting pairing.
Conventional RPA/FLEX studies predict an $s_{\pm}$-wave state driven by spin fluctuations
\cite{T1-Sakakibara,T3-Zhang,T4-Zhang,T5-Zhang}.
Here, using the BS-equation method~\cite{T14-Yamakawa}, we show that the CDW-mediated interaction strongly enhances the superconducting transition temperature of the $s_{\pm}$-wave state.
Figure~\ref{fig5}(a) schematically illustrates the CDW- and SDW-mediated pairing processes at ${\bm Q}_{\rm cdw}$ and ${\bm Q}_{\rm sdw}$, respectively.

The singlet pairing interaction is written as
\begin{equation}
\hat V^{\rm SC}_{\rm BS-eq.}
=
\hat V^{{\rm BS},c} + \hat V^{{\rm BS},s} + \frac{1}{2} (\hat C^s-\hat C^c),
\end{equation}
where the charge- and spin-channel contributions are obtained from the full four-point vertices of the BS equation:
\begin{align}
\hat V^{{\rm BS},c}(k,k')
&=
-\frac{1}{2}
\hat\Gamma^c_{{\bm k}'-{\bm k}}(k,-k') +\frac{1}{2}\hat C^c,
\\
\hat V^{{\rm BS},s}(k,k')
&=
\frac{3}{2}
\hat\Gamma^s_{{\bm k}'-{\bm k}}(k,-k') -\frac{3}{2}\hat C^s,
\end{align}
in which the first-order terms in ${\hat C}^x$ are cancelled.
Near the CDW instability, the charge-channel vertex provides a strong attractive pairing contribution associated with bond-order fluctuations.
This contribution is absent at comparable strength in the conventional FLEX interaction, $\hat V^{\rm SC}_{\rm FLEX}=\hat V^c+\hat V^s+({\hat C}^s-{\hat C}^c)/2$, because FLEX does not capture the vertex-correction-driven enhancement of the charge channel.
For the numerical comparison, we approximate $\hat V^{{\rm BS},s}\approx\hat V^s$ \cite{T14-Yamakawa} and use $\hat V^{\rm SC}_{\rm BS-eq.}\approx \hat V^{{\rm BS},c}+\hat V^s+({\hat C}^s-{\hat C}^c)/2$.
The same dressed Green functions and spin-mediated interaction are thus used in both calculations, isolating the effect of the enhanced charge-channel pairing interaction.
Details are given in Sec.~\ref{sec:methodD} and Appendix~\ref{app:G}.

Figures~\ref{fig5}(b) and \ref{fig5}(d) compare the normalized gap functions obtained from the BS and FLEX interactions for $\Delta E=0$ and $0.1$, respectively.
Both approaches yield essentially the same $s_{\pm}$-wave gap structure, with red and blue indicating opposite signs of the gap.
In both methods, $\Delta_x>0$ for $x=\alpha,\gamma$ and $\Delta_y<0$ for $y=\alpha',\beta',\beta$, while $|\Delta_{\beta}|$ is smaller than the others.
Their pairing strengths, however, differ substantially: the CDW-mediated attraction strongly enhances the superconducting eigenvalue $\lambda_{\rm SC}$.
Figures~\ref{fig5}(c) and \ref{fig5}(e) show the temperature dependences of $\lambda_{\rm SC}$ obtained from the BS and FLEX interactions.
For both values of $\Delta E$, the additional CDW fluctuations enhance the calculated superconducting transition temperature to the 100-K scale while preserving the gap symmetry favored by spin fluctuations.

The origin of this cooperation lies in the $M_z$-parity selection rules illustrated in
Fig.~\ref{fig5}(a).
To interpret the pairing interaction, we focus on the leading mode in each channel.
This approximation is justified near the corresponding instability, provided that subleading modes remain noncritical.
It is used only for the physical interpretation; the numerical BS calculation retains the full vertex without projection onto the leading mode.
Near a density-wave instability, the band-resolved pairing interaction in channel $x=c,s$ takes the approximate form
\begin{equation}
V^{{\rm BS},x}_{ab}({\bm k},{\bm k}')
\approx
|g^x_{ab}({\bm k}';{\bm q})|^2
\mathcal V^x({\bm q}),
\qquad
{\bm q}={\bm k}-{\bm k}',
\end{equation}
with
\begin{equation}
\mathcal V^x({\bm q})
\propto
\frac{v^x}{1-\lambda^x({\bm q})},
\end{equation}
where $a,b$ are band indices, $g^x_{ab}$ is the form factor in the band basis, and $v^c<0$ and $v^s>0$.
The form-factor matrix elements therefore determine which bands are coupled by the enhanced fluctuations.

As established above, the leading CDW form factor is even under $M_z$, whereas the leading SDW form factor is odd.
In the reduced OL subspace of the dominant $d_{3z^2-r^2}$ sector, their nonzero matrix elements in the mirror-parity basis are
\begin{align}
g^c_{++}
&=\delta\varepsilon+\delta t,
&
g^c_{--}
&=\delta\varepsilon-\delta t,
\\
g^s_{+-}
&=g^s_{-+}=\delta m.
\end{align}
Here, $\delta t$, $\delta\varepsilon$, and $\delta m$ denote the inter-OL bond, OL local-potential, and antiphase OL spin modulations, respectively.
The derivation and the reduced-subspace approximation are described in Appendix~\ref{app:C}.
The even CDW mode couples states of the same mirror parity, whereas the odd SDW mode couples states of opposite mirror parities.

The calculated gap structure is consistent with the complementary scattering channels resolved in the BS interaction calculated in Appendix~\ref{app:G}: the dominant CDW-mediated attraction connects same-parity pockets with the same gap sign, whereas the dominant SDW-mediated repulsion connects opposite-parity pockets with opposite gap signs. These parity-selective interactions therefore cooperate to stabilize the $s_{\pm}$-wave state and enhance its transition temperature to the 100-K scale in our calculations.

Cooperative pairing mediated by charge and spin fluctuations has also been discussed in Fe-based superconductors~\cite{T28-Kontani,T15-Kontani}.
In the present trilayer system, mirror-parity selection rules provide a microscopic basis for this cooperation.

\subsection{Comparison with bilayer nickelates}

To examine the applicability of the present mechanism beyond the trilayer system, we also analyze a bilayer La$_3$Ni$_2$O$_7$ model within the same framework, including the FLEX self-energy consistently (see Appendix~\ref{app:B}).
Quantum interference generates a strong CDW instability at ${\bm q}\approx(\pi/2,\pi/2)$, dominated by interlayer bond order.
\textcolor{black}{Resonant x-ray scattering on bilayer films reports magnetic order and charge-like anisotropy with the same periodicity~\cite{E12-Ren}.}
The deviation from ${\bm Q}_{\rm cdw}\approx2{\bm Q}_{\rm sdw}$ reflects the additional role of Fermi-surface nesting in selecting the ordering wavevector within the momentum range enhanced by paramagnon interference.

The BS-equation analysis further yields an $s_{\pm}$-wave superconducting state with $T_c\sim100$~K through cooperative CDW and SDW fluctuations.
\textcolor{black}{As in the trilayer system, the leading CDW and SDW form factors
have $M_z$ parities of $+1$ and $-1$, respectively, allowing
their fluctuations to cooperate efficiently in stabilizing
$s_{\pm}$-wave superconductivity through mirror-parity
selection rules.} 
A related enhancement of the $s_{\pm}$-wave pairing by out-of-plane electron--phonon coupling was reported in an FRG study~\cite{T30-Zhan}.
In the present theory, the attraction arises from electronically generated bond-order fluctuations.
Thus, neither the bond-order instability nor the enhanced superconductivity requires the additional $\alpha'$ pocket of the trilayer system.
These results extend the quantum-interference and cooperative pairing mechanisms to both bilayer and trilayer nickelates.

\section{Summary}

We have identified a microscopic mechanism linking intertwined density-wave order and high-temperature superconductivity in trilayer La$_4$Ni$_3$O$_{10}$.
Quantum interference between paramagnons enhances the charge channel at ${\bm Q}_{\rm cdw}\approx2{\bm Q}_{\rm sdw}$, providing a microscopic basis for the experimentally observed CDW--SDW wavevector relation.

The resulting order has a distinctive layer structure: paramagnons on the OLs generate an inter-OL bond order, which unexpectedly induces the strongest orbital polarization on the IL.
The Ni $d_{3z^2-r^2}$ and $d_{x^2-y^2}$ occupations modulate predominantly in antiphase, producing a large orbital polarization with only weak total-charge modulation.
This intertwined bond-and-orbital order accounts for the layer-selective electronic reconstruction inferred from NMR/NQR and is consistent with Raman, STM, and diffraction measurements.

The same density-wave fluctuations also cooperate in superconducting pairing.
The $M_z$-parity selection rules direct the CDW-mediated attraction and SDW-mediated repulsion into complementary scattering channels that support the same $s_{\pm}$-wave gap structure, yielding a calculated $T_c$ on the 100-K scale.
The corresponding results for the bilayer system demonstrate that this mechanism extends beyond trilayer nickelates.
Together, these findings identify quantum interference and mirror-parity selection as complementary principles linking density-wave formation to cooperative pairing in multilayer nickelates.

\section{Methods}\label{sec:methods} 

\subsection{Construction of the trilayer multiorbital model from first principles}
\label{sec:methodA}

We construct a six-band tight-binding model for trilayer La$_4$Ni$_3$O$_{10}$ consisting of the Ni $d_{3z^2-r^2}$ and $d_{x^2-y^2}$ orbitals on each of the three NiO$_2$ layers.
The electronic structure is obtained from first-principles calculations using the WIEN2k package~\textcolor{black}{\cite{T34-Blaha}} and downfolded onto the Ni $e_g$ Wannier basis using Wannier90~\textcolor{black}{\cite{T35-Pizzi}}.
We use the experimentally determined tetragonal $I4/mmm$ crystal structure at $P=16.6$~GPa \cite{E11-Li}.

First-principles calculations were performed using the full-potential linearized augmented-plane-wave plus local-orbital (FP-LAPW+lo) method implemented in WIEN2k. The exchange-correlation functional was treated within the generalized gradient approximation (GGA) of Perdew, Burke, and Ernzerhof (PBE). We used $R_{\rm MT}K_{\rm max}=7.0$ and a $10 \times 10 \times 10$ k-point mesh.

We introduce a composite orbital--layer index $m=(d,j)$, where
$d=d_{3z^2-r^2},d_{x^2-y^2}$ and $j=1,2,3$.
The two outer NiO$_2$ layers correspond to $j=1,3$, whereas
$j=2$ denotes the inner layer.
To examine the sensitivity of the results to the position of the
$d_{3z^2-r^2}$-dominated $\gamma$ band around the M point, we
introduce an energy shift $\Delta E$.
The explicit form of this shift and the resulting evolution of the
$\gamma$ pocket are given in Appendix~\ref{app:A}.

In the present model, ($n_{3z^2-r^2}, \ n_{x^2-y^2})=(0.65,0.55)$ for the OL, and $(0.80,0.60)$ for the IL.
The spin degeneracy $2$ is taken into account.

\subsection{Local Coulomb interactions and conserving approximation}
\label{sec:methodB}

The local Coulomb interaction is parametrized by the intraorbital repulsion $U$, interorbital repulsion $U'$, Hund's coupling $J$, and pair-hopping interaction $J'$.
We impose the rotationally invariant relations $U'=U-2J$ and $J'=J$, and set $U=3$ and $J/U=0.1$ unless otherwise stated.

Our theoretical framework is based on the conserving approximation of Baym and Kadanoff~\textcolor{black}{\cite{T32-BaymKadanoff,T33-Baym}}.
Specifically, the FLEX approximation~\textcolor{black}{\cite{T29-Bickers}} is formulated in terms of a Luttinger--Ward (LW) functional~\textcolor{black}{\cite{T31-LuttingerWard}}
$\Phi_{\rm LW}[G]$, from which the self-energy is obtained as
$\Sigma=\delta\Phi_{\rm LW}/\delta G$
(see Appendix~\ref{app:E} for explicit expressions).
In the conventional FLEX treatment of the present model with $J>0$, the CDW instability is weaker than the SDW instability.
However, a conserving evaluation of the susceptibilities requires vertex corrections consistent with the self-energy.
We incorporate these corrections through the DW
equation and show that CDW order can precede SDW order.

In the present study, we use a momentum mesh of $N_k=64\times64$ points and $N_\omega=8192$ Matsubara frequencies for the FLEX the DW equation analyses.

\subsection{Density-wave equation}
\label{sec:methodC}

Conventional charge and spin orders have local order parameters defined on atomic sites.
In contrast, unconventional charge orders, such as bond order and loop-current order, have nonlocal order parameters extending between sites.
The DW equation provides a powerful tool for investigating such unconventional orders.

The DW equation follows from the stationarity condition of the LW free-energy functional within the adopted conserving approximation~\cite{T12-Tazai,T14-Yamakawa}.
Without assuming a specific ordering pattern, its leading eigenmode determines the form factor of the incipient density-wave order through the quadratic variation of the free energy.
Introducing pair indices $L=(l,l')$ and $M=(m,m')$, the linearized
charge-channel DW equation is
\begin{equation}
\begin{aligned}
&\lambda_{\bm q} f_{\bm q}^{L}(k)
=-\frac{T}{N}\sum_{k',M_1,M_2}I^{c;L,M_1}_{\bm q}(k,k')
\\
&\qquad\times
\left\{\hat G(k')\hat G(k'+q)\right\}_{M_1,M_2}
f_{\bm q}^{M_2}(k').
\end{aligned}
\end{equation}
Here, $\lambda_{\bm q}$ and $f_{\bm q}^{L}(k)$ are the DW
eigenvalue and form factor, respectively, and the density-wave
transition occurs when the leading eigenvalue reaches unity.

The irreducible kernel consists of the Hartree--Fock, Maki--Thompson, and two
Aslamazov--Larkin contributions.
The Aslamazov--Larkin terms describe quantum interference between
collective fluctuations and provide the dominant contribution to
the finite-${\bm q}$ CDW instability found in the present study.
Explicit expressions for these vertex corrections are given in
Appendix~\ref{app:F}.

The local orbital- and layer-resolved density modulations induced by
the DW form factor are evaluated within linear response using the
FLEX Green functions.
For the results shown in the main text, we set
${\bm q}={\bm Q}_{\rm cdw}$ and normalize the maximum amplitude of
the DW form factor to $f_{\rm max}=0.1$.
The charge and orbital responses are obtained from the sum and
difference, respectively, of the $d_{3z^2-r^2}$ and $d_{x^2-y^2}$
density modulations on each NiO$_2$ layer.

\subsection{Superconducting gap equation and Bethe--Salpeter analysis}
\label{sec:methodD}

The superconducting instability is analyzed by solving the
linearized singlet gap equation in the band basis:
\begin{equation}
\lambda_{\rm SC}\bar{\Delta}_{b}(k)
=
-\frac{T}{N}\sum_{k',b'}
V^{\rm SC}_{bb'}(k,k')
|G_{b'}(k')|^2
\bar{\Delta}_{b'}(k'),
\label{eqn:SC-gap-equation}
\end{equation}
where $b$ and $b'$ denote band indices.
$V^{\rm SC}_{bb'}(k,k')$ is derived from ${\hat V}^{\rm SC}(k,k')$ using the unitary matrix $u_{m,b}({\bm k})$ that transforms the orbital--layer basis to the band basis.
The superconducting transition occurs when $\lambda_{\rm SC}=1$.

The pairing interaction $V^{\rm SC}_{bb'}(k,k')$ contains both a repulsive spin-fluctuation contribution obtained within FLEX and an attractive charge-channel contribution enhanced by the vertex corrections responsible for the CDW instability.
To evaluate the latter beyond the conventional Migdal approximation, we solve the Bethe--Salpeter equation for the full charge-channel four-point vertex and construct the corresponding bond-fluctuation pairing interaction \cite{T14-Yamakawa}.

In trilayer nickelates, the DOS at the Fermi level on the five Fermi pockets are \[
\begin{aligned}
&\bigl(N_{\alpha}(0),N_{\alpha'}(0),N_{\beta}(0),N_{\beta'}(0),N_{\gamma}(0)\bigr)
\\
&\qquad=(0.20,0.25,0.41,0.41,1.92)~\mathrm{eV}^{-1}
\end{aligned}
\] for $\Delta E=0$, whereas $N_{\gamma}(0)$ decreases by approximately 40\% for $\Delta E=0.1$.
The DOS varies rapidly with energy near the Fermi level, particularly for the $\gamma$ pocket.
Therefore, approximating the $\k'$-sum in Eq.~(\ref{eqn:SC-gap-equation}) by a line integral over the Fermi surface is not well justified.
In this work, we evaluate the $\k'$-sum directly on a momentum-space mesh.
The pairing interaction is evaluated on the Fermi surfaces and extended to off-Fermi-surface states within an energy cutoff, as described in Appendix~\ref{app:G}.
Importantly, the internal momentum and frequency sums over Green functions in the BS equation are evaluated on an $N_k$ momentum and $N_\omega$ Matsubara mesh covering the entire Brillouin zone, rather than being restricted to the Fermi surfaces \cite{T14-Yamakawa}.

\begin{acknowledgments}
We are grateful to J. Zhan and S. Onari for fruitful discussions.
\end{acknowledgments}

\appendix

\section{Introduction of the energy shift $\Delta E$ around the M point}
\label{app:A}

The tight-binding model for the trilayer nickelate is introduced in Sec.~\ref{sec:methodA}.
The kinetic Hamiltonian in momentum space is given by
\begin{equation}
H=
\sum_{{\bm k},l,m,\sigma}
h_{{\bm k},l,m}\,
c^{\dagger}_{{\bm k},l\sigma}
c_{{\bm k},m\sigma},
\label{eq:SI-H0}
\end{equation}
where $h_{{\bm k},l,m}$ is the Fourier transform of the real-space tight-binding matrix elements, and $c_{{\bm k},m\sigma}$ annihilates an electron with momentum ${\bm k}$, spin $\sigma$, and index $m$.
The composite indices $l,m=(d,j)$ label the two Ni $e_g$ orbitals, $d=d_{3z^2-r^2},d_{x^2-y^2}$, on the three NiO$_2$ layers, $j=1,2,3$.

Experimentally, the size and even the presence of the $\gamma$ pocket show substantial sample dependence.
To examine the sensitivity to the $\gamma$ pocket, we introduce an energy-shift parameter $\Delta E$ for the $d_{3z^2-r^2}$-dominated band around the M point.
The explicit functional form of this shift is
\begin{equation}
\Delta h_{{\bm k},l,m}=-\frac{\Delta E}{2}\left(\cos k_x+\cos k_y\right)\delta_{j_l,j_m}\delta_{d_l,1}\delta_{d_m,1}.
\end{equation}
This parameter allows us to continuously vary the size of the $\gamma$ pocket and, for sufficiently negative $\Delta E$, remove it from the Fermi surface.

The evolution of the $\gamma$-band dispersion with $\Delta E$ is illustrated in the inset of Fig.~\ref{fig1}(c) of the main text, while the electronic structure for $\Delta E=-0.1$, where the $\gamma$ pocket is absent, is shown in Fig.~\ref{fig11}(a).

\section{Results for a bilayer nickelate model for La$_3$Ni$_2$O$_7$}
\label{app:B}

\begin{figure*}[!tp]
\centering
\includegraphics[width=0.85\textwidth]{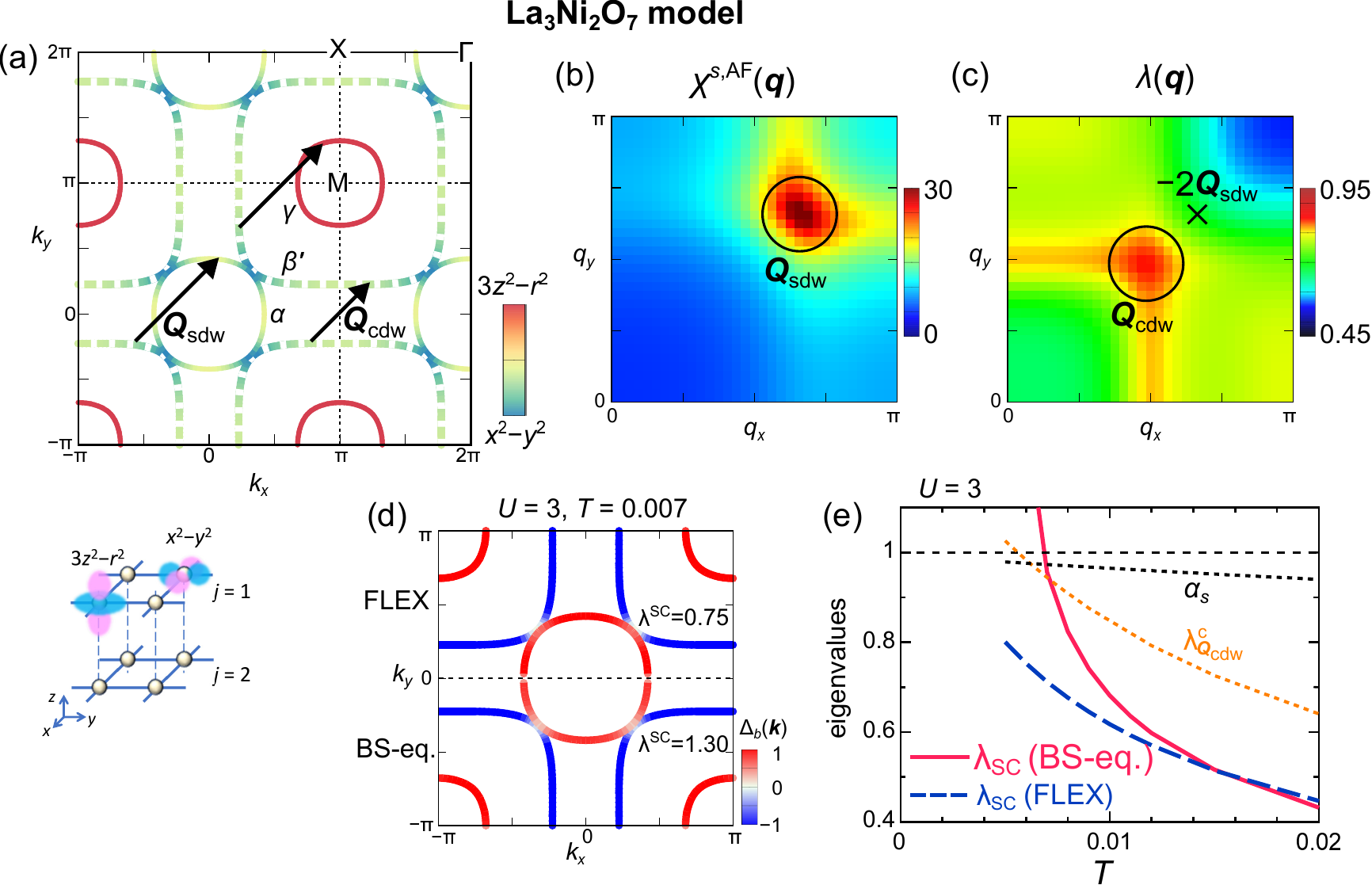}
\caption{
{\bf Density-wave correlations and superconductivity in a bilayer nickelate model:}  \ 
(a) Fermi surfaces ($\a$,$\b'$,$\gamma$) and orbital weights.
The bilayer lattice structure is also illustrated.
(b) Dominant interlayer-antiferromagnetic spin susceptibility obtained by FLEX.
(c) Charge-channel DW eigenvalue generated by the quantum-interference mechanism.
(d) $s_{\pm}$-wave gap structures obtained from the SDW-fluctuation mechanism (FLEX) and the combined CDW+SDW fluctuation mechanism (BS equation).
(e) Temperature dependence of $\lambda_{\rm SC}$, $\lambda_{{\rm Q}_{\rm cdw}}$, and $\a_S$.
}
\label{fig6}
\end{figure*}

In Ref.~\cite{T13-Inoue}, we analyzed the bilayer nickelate model by neglecting the electronic self-energy and retaining interactions only for the $d_{3z^2-r^2}$ orbital.
Here, we extend this analysis to include self-energy corrections and interactions involving both Ni $e_g$ orbitals, with the results summarized in Fig.~\ref{fig6}.
The present approach fulfills the requirements of a conserving approximation, providing a more consistent theoretical basis for the numerical results.
It also yields robust results without requiring fine-tuning of the model or its parameters.

Figure~\ref{fig6}(a) shows the Fermi surfaces and their orbital characters in the bilayer model.
Compared with the trilayer system, the $\alpha'$ and $\beta$ Fermi-surface sheets are absent.
Figure~\ref{fig6}(b) shows the momentum dependence of the spin fluctuations obtained by FLEX.
We plot the dominant interlayer antiferromagnetic component, which develops a pronounced peak around
${\bm q}\approx(0.65\pi,0.65\pi)$.
Figure~\ref{fig6}(c) shows the CDW eigenvalue obtained from the DW equation.
The CDW fluctuations are strongly enhanced by the quantum-interference mechanism and exhibit a peak at
${\bm q}\approx(\pi/2,\pi/2)$.
\textcolor{black}{Resonant x-ray scattering on bilayer films reveals a reflection at $(1/4,1/4,L)$ associated with magnetic order and charge-like anisotropy~\cite{E12-Ren}, providing an experimental comparison for the calculated in-plane modulation.}
These results are consistent with the reported coexistence of SDW and CDW correlations in the bilayer nickelate~\textcolor{black}{\cite{E15-Luo}}.

The quantum-interference-driven CDW is particularly enhanced when the interference wavevector
${\bm q}_{\rm qi}\equiv{\bm Q}_{\rm sdw}\pm{\bm Q}^{(\prime)}_{\rm sdw}$
coincides with a favorable Fermi-surface nesting wavevector ${\bm q}_{\rm nest}$.
This condition is realized in La$_4$Ni$_3$O$_{10}$.
When ${\bm q}_{\rm qi}$ and ${\bm q}_{\rm nest}$ do not coincide exactly, the actual CDW wavevector tends to shift toward a nesting wavevector ${\bm q}_{\rm nest}$ close to ${\bm q}_{\rm qi}$, because the spin fluctuations above $T_{\rm SDW}$ have a finite width in momentum space.
In La$_3$Ni$_2$O$_7$, the CDW wavevector is shifted toward ${\bm q}_{\rm nest}$, so that the relation ${\bm Q}_{\rm cdw}=2{\bm Q}_{\rm sdw}$ is not exactly satisfied.

When $T_{\rm SDW}<T_{\rm CDW}$, coupling to the pre-existing CDW may favor an adjustment of the SDW wavevector toward $2{\bm Q}_{\rm sdw}={\bm Q}_{\rm cdw}$, modulo a reciprocal-lattice vector.
Assuming this wavevector relation, a symmetry-allowed coupling has the form
$\Delta F\sim {\rm Re}\{M_{{\bm Q}_{\rm sdw}}^2\rho_{-2{\bm Q}_{\rm sdw}}e^{i\phi_0}\}$.
Here, $M$ and $\rho$ denote the SDW and CDW mode amplitudes, with their relative phase written explicitly as $\phi_0$.
This coupling provides a possible energetic mechanism for wavevector adjustment, which is not calculated here.

We next solve the superconducting gap equation.
As shown in Fig.~\ref{fig6}(d), both the spin-fluctuation mechanism treated within FLEX and the combined CDW+SDW fluctuation mechanism treated by the BS-equation method yield an $s_{\pm}$-wave superconducting state.
In both methods, $\Delta_x>0$ for $x=\alpha,\gamma$ and $\Delta_y<0$ for $y=\beta'$.
Figure~\ref{fig6}(e) presents the temperature dependence of the superconducting eigenvalue.
For the combined CDW+SDW fluctuation mechanism, the eigenvalue reaches unity at a temperature on the 100~K scale.
\textcolor{black}{As in the trilayer system, the leading CDW (SDW) form factor
has $M_z$ parity $+1$ ($-1$).
Accordingly, SDW fluctuations predominantly mediate repulsive
scattering between opposite-parity pockets, mainly
$\alpha$--$\beta'$ and $\beta'$--$\gamma$, whereas CDW
fluctuations predominantly mediate attractive scattering
within the same parity sector, mainly within the $\beta'$ pocket.
These complementary interactions support cooperative
$s_{\pm}$-wave pairing.}
For the bilayer model, the DW equation yields $\delta t\sim-\delta\varepsilon$, so that $|g^c_{--}|>|g^c_{++}|$ and the dominant-mode contribution satisfies $|V^c_{--}|>|V^c_{++}|$.
This is opposite to the relative strengths in the trilayer model and explains the different dominant intraband attractive channels.
\textcolor{black}{The mirror-parity selection rules are derived in
Appendix~\ref{app:C}.}

\section{$M_z$-parity selection rule for CDW- and SDW-fluctuation-mediated pairing}
\label{app:C}

We interpret the pairing interaction in terms of the leading mode in each channel $x=c,s$.
This approximation is justified near the corresponding instability when the leading eigenvalue approaches unity while subleading modes remain noncritical.
Subleading modes can have different mirror parities and contribute additional scattering channels.
The numerical BS calculation retains the full vertex without projection onto a single mode; the approximation below is used only to explain its dominant structure.

For the in-plane momenta considered here, the layer-mirror operation $M_z$ leaves momentum unchanged.
Let $p_a,p_b=\pm1$ be the mirror eigenvalues of the electronic states and $\eta_x=\pm1$ the parity of a given mode, defined by $M_z\hat f^x M_z^{-1}=\eta_x\hat f^x$.
The band-basis matrix element is
\[
g^x_{ab}(k;q)=\sum_{m,l}u^*_{m,a}({\bm k}+{\bm q})f^x_{ml}(k;q)u_{l,b}({\bm k}),
\]
where $m,l$ include both orbital and layer indices.
Mirror symmetry gives $g^x_{ab}=\eta_xp_ap_b g^x_{ab}$, or equivalently
\[
g^x_{ab}=0\qquad\text{unless}\qquad p_ap_b=\eta_x.
\]
This selection rule is exact for each mode and does not require an outer-layer approximation.
The leading CDW mode has $\eta_c=+1$ and couples states of the same parity, whereas the leading SDW mode has $\eta_s=-1$ and couples states of opposite parities.

To obtain simple expressions for the amplitudes, we retain the dominant $d_{3z^2-r^2}$ components in the OL subspace $(|j=1\rangle,|j=3\rangle)$.
The reduced form factors are $\hat f^c=\delta\varepsilon\hat\sigma_0+\delta t\hat\sigma_x$ and $\hat f^s=\delta m\hat\sigma_z$.
The parity states are $(|j=1\rangle\pm|j=3\rangle)/\sqrt{2}$, and $\hat u=(\hat\sigma_z+\hat\sigma_x)/\sqrt{2}$ transforms these matrices to the parity basis.
Thus, $\hat g^x=\hat u^\dagger\hat f^x\hat u$ gives
\[
g^c_{++}=\delta\varepsilon+\delta t,\qquad
g^c_{--}=\delta\varepsilon-\delta t,\qquad
g^s_{+-}=g^s_{-+}=\delta m.
\]
These reduced-subspace amplitudes omit the detailed orbital and layer weights of the band eigenstates; the full band transformation is used in the numerical calculation.

Within this leading-mode description, the interaction strengths are proportional to $|g^x_{ab}|^2\mathcal V^x(q)$, with $\mathcal V^c<0$ and $\mathcal V^s>0$.
The resulting dominant intraparity attraction and interparity repulsion are resolved in the numerical BS interaction in Appendix~\ref{app:G}.
Their compatibility with the calculated gap signs explains the cooperative enhancement of $s_{\pm}$-wave pairing without requiring a uniform gap sign throughout each parity sector.
Related cooperative charge- and spin-fluctuation pairing has also been discussed in iron-based superconductors~\cite{T28-Kontani,T15-Kontani}.

\section{Orbital and charge polarization induced by SDW order}
\label{app:D}

The primary bond-order instability produces an IL-centered orbital response above the SDW transition, as discussed in the main text.
Below $T_{\rm SDW}$, NMR/NQR measurements suggest that charge-related modulations also develop on the OLs \cite{E5-Wang}.
A possible additional contribution is the charge and orbital response induced at second order in the SDW order parameter.
This response occurs at $2{\bm Q}_{\rm sdw}$, which coincides with the primary density-wave wavevector up to its sign and a reciprocal-lattice vector.
It may therefore modify the layer distribution of the density modulation below $T_{\rm SDW}$.
Determining its amplitude and interlayer phase structure requires a calculation in the SDW-ordered state and is left for future work.

\section{FLEX self-energy and quasiparticle mass enhancement}
\label{app:E}

\begin{figure}[!tbp]
\centering
\includegraphics[width=\linewidth]{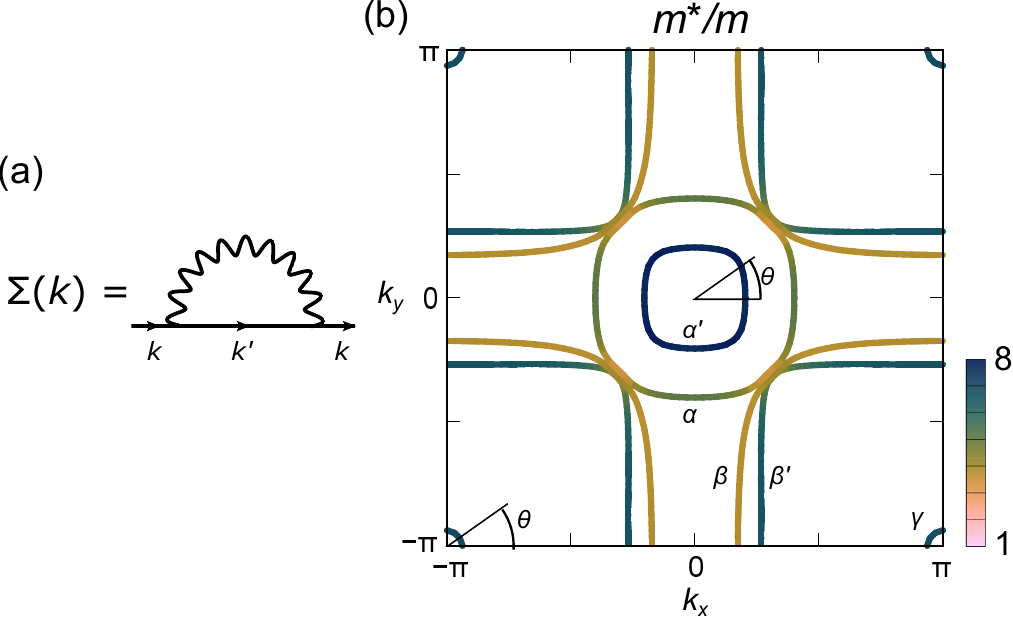}
\caption{
{\bf FLEX self-energy and quasiparticle mass enhancement:} \ 
(a) Self-energy diagram in the FLEX approximation.
(b) Quasiparticle mass-enhancement factor on the Fermi surfaces obtained from the FLEX self-energy for $U=3$ and $T=0.01$.
}
\label{fig8}
\end{figure}

Figure~\ref{fig8}(a) shows the self-energy diagram included in the multiorbital FLEX approximation
\cite{T29-Bickers}.
For completeness, we summarize here the FLEX formalism used in the present calculation.

For $U=0$, the noninteracting Green function is
\begin{equation}
\hat G_0(k)
=
\left[
(i\epsilon_n+\mu)\hat 1-\hat h_{\bm k}
\right]^{-1},
\label{eq:SI-G0}
\end{equation}
where $k=({\bm k},\epsilon_n)$, $\epsilon_n=(2n+1)\pi T$, and $\mu$ is the chemical potential, retained explicitly in the Green function.
When interaction effects are included, the dressed Green function satisfies the Dyson equation
\begin{equation}
[\hat G(k)]^{-1}
=
[\hat G_0(k)]^{-1}
-
\hat\Sigma(k),
\label{eq:SI-Dyson}
\end{equation}
where $\hat\Sigma(k)$ is the electronic self-energy.

In the FLEX approximation, the self-energy is given by the convolution of ${\hat G}$ and the effective interaction $\hat V$, 
which is constructed from the spin and charge susceptibilities.
It is expressed as
\begin{align}
\Sigma_{lm}(k)
&=
\frac{T}{N}\sum_{q}\sum_{l'm'}
V_{ll',mm'}(q)G_{l'm'}(k-q),
\label{eq:SI-FLEX-Sigma}
\\
\hat V(q)
&=
\frac{3}{2}\hat C^s\hat\chi^s(q)\hat C^s
+\frac{1}{2}\hat C^c\hat\chi^c(q)\hat C^c
\nonumber\\*
&\hspace{2cm}
-\frac{3}{4}\hat C^s\hat\chi^0(q)\hat C^s
-\frac{1}{4}\hat C^c\hat\chi^0(q)\hat C^c,
\label{eq:SI-FLEX-V}
\\
\hat\chi^{s(c)}(q)
&=
\hat\chi^0(q)
\left[
\hat 1-\hat C^{s(c)}\hat\chi^0(q)
\right]^{-1},
\label{eq:SI-FLEX-chi}
\\
\chi^0_{ll',mm'}(q)
&=
-\frac{T}{N}\sum_k
G_{lm}(k+q)G_{m'l'}(k),
\label{eq:SI-FLEX-chi0}
\end{align}
where $k=({\bm k},\epsilon_n)$ with $\epsilon_n=(2n+1)\pi T$ and
$q=({\bm q},\omega_\nu)$ with $\omega_\nu=2\nu\pi T$.
The indices $l,m,\ldots$ in Eqs.~(\ref{eq:SI-FLEX-Sigma})--(\ref{eq:SI-FLEX-chi0})
denote the six orbital--layer basis states of the trilayer two-orbital model.

At each FLEX iteration, we subtract the static Hermitian part of the self-energy, $\Delta{\hat\Sigma}_\k=[\hat\Sigma^{\rm R}(\k,0)+\hat\Sigma^{\rm A}(\k,0)]/2$, to preserve the original Fermi surfaces.
This term is nearly diagonal in the orbital basis and approximately $\k$-independent, corresponding to small shifts in the orbital energies on each layer.

The local bare Coulomb vertices act only within the same NiO$_2$ layer.
For a fixed NiO$_2$ layer $j$, the nonzero local Coulomb matrix elements are specified by the orbital indices $d_1,d_2,d_3,d_4$.  The spin-channel vertex is
\begin{equation}
(C^s)_{d_1d_2,d_3d_4}
=
\begin{cases}
U, & d_1=d_2=d_3=d_4,\\
U', & d_1=d_3\neq d_2=d_4,\\
J, & d_1=d_2\neq d_3=d_4,\\
J, & d_1=d_4\neq d_2=d_3,\\
0, & \mathrm{otherwise},
\end{cases}
\label{eq:SI-Cs}
\end{equation}
where $U'=U-2J$.
The corresponding charge-channel vertex is
\begin{equation}
(C^c)_{d_1d_2,d_3d_4}
=
\begin{cases}
-U, & d_1=d_2=d_3=d_4,\\
U'-2J, & d_1=d_3\neq d_2=d_4,\\
-2U'+J, & d_1=d_2\neq d_3=d_4,\\
-J, & d_1=d_4\neq d_2=d_3,\\
0, & \mathrm{otherwise}.
\end{cases}
\label{eq:SI-Cc}
\end{equation}
The corresponding matrix elements in the six-state composite basis additionally contain the condition that all four indices belong to the same layer $j$; matrix elements connecting different NiO$_2$ layers vanish because the bare interaction is local.

The self-consistent FLEX equations incorporate both the quasiparticle damping and the orbital- and momentum-dependent band renormalization generated by the spin and charge fluctuations.
Using the resulting self-energy, we evaluate the quasiparticle mass-enhancement factor on the Fermi surfaces.
First, using Pade analytic continuation, the Matsubara self-energy $\Sigma_{lm}({\bm k},i\epsilon_n)$ is analytically continued from the upper half of the complex-frequency plane to the real-energy axis as
\begin{equation}
\Sigma_{lm}^{R}({\bm k},\epsilon)
=
\left.
\Sigma_{lm}({\bm k},z)
\right|_{z\to\epsilon+i0^{+}},
\label{eq:SI-SigmaR}
\end{equation}
which gives the retarded self-energy.
Using the unitary matrix $u_{m,b}({\bm k})$ that transforms
the orbital--layer basis to the band basis, we obtain
the diagonal band self-energy as
\begin{equation}
\Sigma_b^{R}({\bm k},\epsilon)
=
\sum_{l,m}
u^{*}_{l,b}({\bm k})
\Sigma_{lm}^{R}({\bm k},\epsilon)
u_{m,b}({\bm k}).
\label{eq:SI-Sigma-band}
\end{equation}
For a momentum ${\bm k}$ on the Fermi surface of band $b$, we define the quasiparticle mass-enhancement factor by
\begin{equation}
Z_b({\bm k})
=
{\rm Re}
\left\{
1-
\left.
\frac{\partial \Sigma_b^{R}({\bm k},\epsilon)}
{\partial\epsilon}
\right|_{\epsilon=0}
\right\}.
\label{eq:SI-mass-enhancement}
\end{equation}
The momentum dependence of $Z_b({\bm k})$ obtained for $U=3$ and $T=0.01$ is shown in Fig.~\ref{fig8}(b).
These self-energy effects are retained consistently in the subsequent DW- and BS-equation analyses.

\section{CDW form factor obtained from the DW equation}
\label{app:F}

\begin{figure*}[!tp]
\centering
\includegraphics[width=0.70\textwidth]{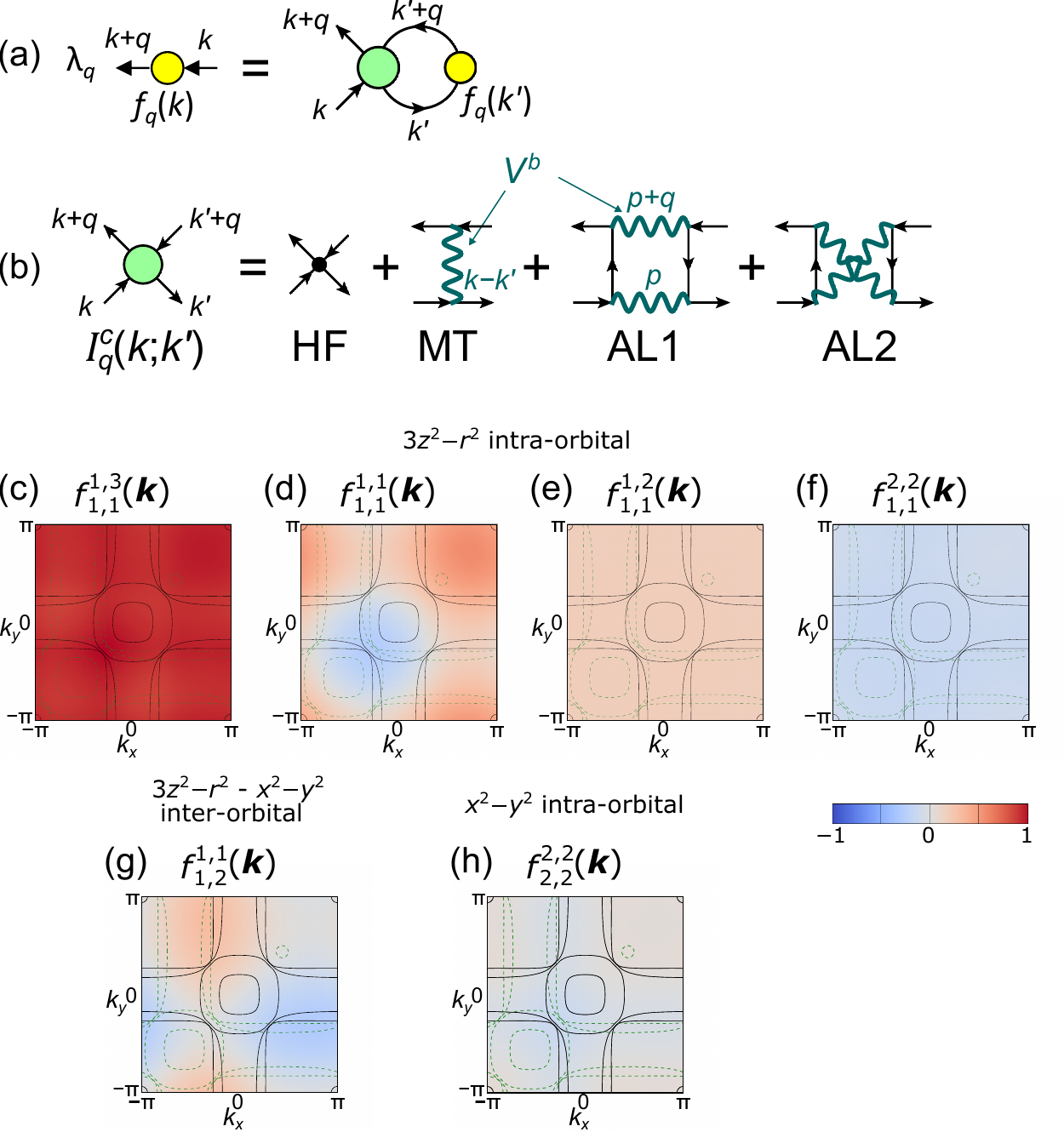}
\caption{
{\bf Density-wave equation and microscopic CDW form factor:} \ 
(a) Diagrammatic representation of the charge-channel DW equation.
(b) Irreducible kernel $I^c$, consisting of the HF, MT, AL1, and AL2 terms.
(c)--(h) Representative components of the CDW form factor at $\q={\bm Q}_{\rm cdw}$.
Here, the form factor is normalized as $\max_{l,m,\k} |f^{lm}_{\bm q}({\bm k})|=1$.
The dominant $f^{1,3}_{1,1}(\k)$ component corresponds to the interlayer bond-order modulation $\delta t$.
Here, ${\hat f}_{\bm q}({\bm k})$ is the Hermitian part of ${\hat f}_{\bm q}({\bm k},\e-i0)$ at zero energy.
}
\label{fig9}
\end{figure*}

In Sec.~\ref{sec:methodC}, we have introduced the linearized charge-channel DW equation utilized in the main text.
Using the pair indices $L=(l,l')$ and $M=(m,m')$, it is expressed as
\begin{equation}
\begin{aligned}
&\lambda_{{\bm q}} f_{{\bm q}}^{L}(k)
=-\frac{T}{N}\sum_{k',M_1,M_2}I_{{\bm q}}^{c;L,M_1}(k,k')
\\
&\qquad\times
\left\{\hat G(k')\hat G(k'+q)\right\}_{M_1,M_2}
f_{{\bm q}}^{M_2}(k').
\end{aligned}
\label{eq:SI-DW}
\end{equation}
Here, $f_{{\bm q}}^{L}(k)$ is the charge-channel form factor and $\lambda_{{\bm q}}$ is the corresponding eigenvalue.
The CDW transition occurs when the leading eigenvalue reaches unity.
Equation~(\ref{eq:SI-DW}) describes an electron--hole pairing instability mediated by the irreducible kernel $\hat I_{{\bm q}}^{c}(k,k')$.

The conserving approximation is based on the Luttinger--Ward (LW) functional $\Phi_{\rm LW}[G]$.
The self-energy is obtained as $\Sigma_M(k)=\delta\Phi_{\rm LW}/\delta \hat G_M(k)$.
The charge-channel kernel at $q=0$ is obtained from the second functional derivative of the same LW functional;
${\hat I}^{c}_{M,L}(k,k')=\delta^2\Phi_{\rm LW}/\delta G_M(k)\delta G_L(k')$.
In the one-loop approximation for $\Phi_{\rm LW}$ used here, the charge-channel kernel shown in Fig.~\ref{fig9}(b) is decomposed as \cite{T12-Tazai}
\begin{equation}
\hat I_{{\bm q}}^{c}
=
\hat I_{{\bm q}}^{\rm HF}
+
\hat I_{{\bm q}}^{\rm MT}
+
\hat I_{{\bm q}}^{\rm AL1}
+
\hat I_{{\bm q}}^{\rm AL2},
\label{eq:SI-Idecomp}
\end{equation}
where the four terms are the HF, MT, AL1, and AL2 contributions, respectively.
Their analytic expressions for general $q$ are given as 
\begin{align}
&I_{{\bm q};ll',mm'}^{\rm HF}(k,k')=C^c_{ll',mm'},
\label{eq:SI-IH}
\\
&I_{{\bm q};ll',mm'}^{{\rm MT}}(k,k')
\nonumber\\*
&\quad=-\frac{1}{2}\sum_{b=s,c}a_b
\left[V^b_{lm,l'm'}(k-k')-C^b_{lm,l'm'}\right],
\label{eq:SI-IMT}
\\
&I_{{\bm q};ll',mm'}^{\rm AL1}(k,k')
\nonumber\\*
&\quad=\frac{T}{N}\sum_{b=s,c}\sum_{p,l_1l_2m_1m_2}\frac{a_b}{2}
\nonumber\\*
&\qquad\times V^b_{ll_1,mm_1}(p+q)V^b_{m'm_2,l'l_2}(p)
\nonumber\\*
&\qquad\times G_{l_1l_2}(k-p)G_{m_2m_1}(k'-p),
\label{eq:SI-IAL1}
\\
&I_{{\bm q};ll',mm'}^{{\rm AL2}}(k,k')
\nonumber\\*
&\quad=\frac{T}{N}\sum_{b=s,c}\sum_{p,l_1l_2m_1m_2}\frac{a_b}{2}
\nonumber\\*
&\qquad\times V^b_{ll_1,m_2m'}(p+q)V^b_{m_1m,l'l_2}(p)
\nonumber\\*
&\qquad\times G_{l_1l_2}(k-p)G_{m_2m_1}(k'+p+q).
\label{eq:SI-IAL2}
\end{align}
Here, $a_s=3$, $a_c=1$, and
\begin{equation}
\hat V^b(q)
=
\hat C^b
+
\hat C^b\hat\chi^b(q)\hat C^b
\qquad (b=s,c)
\label{eq:SI-Vb}
\end{equation}
is the fluctuation-mediated interaction in channel $b$.
The double counting of the second-order terms in the Coulomb interaction is subtracted in the numerical implementation.
The HF contribution is first order in the local Coulomb interaction, whereas the AL terms describe the quantum interference between collective fluctuations.
For the present CDW instability, the bond-order form factor is generated predominantly by the two AL terms, 
whereas the HF term is important to suppress the unphysical local charge order.
This explicitly demonstrates that the CDW is not a conventional mean-field charge instability but is driven by the quantum-interference mechanism.

Figures~\ref{fig9}(c)--\ref{fig9}(h) show representative components of the CDW form factor obtained from Eq.~(\ref{eq:SI-DW}).
The largest component is $f^{1,3}_{1,1}$, corresponding to an interlayer bond order on $d_{3z^2-r^2}$ orbital.
Its weak momentum dependence indicates that the dominant real-space component is
a vertical inter-OL bond modulation along the $c$ axis, denoted by $\delta t$ in the main text.

\section{Superconducting interaction from the BS-equation method}
\label{app:G}

\begin{figure*}[!tp]
\centering
\includegraphics[width=0.9\textwidth]{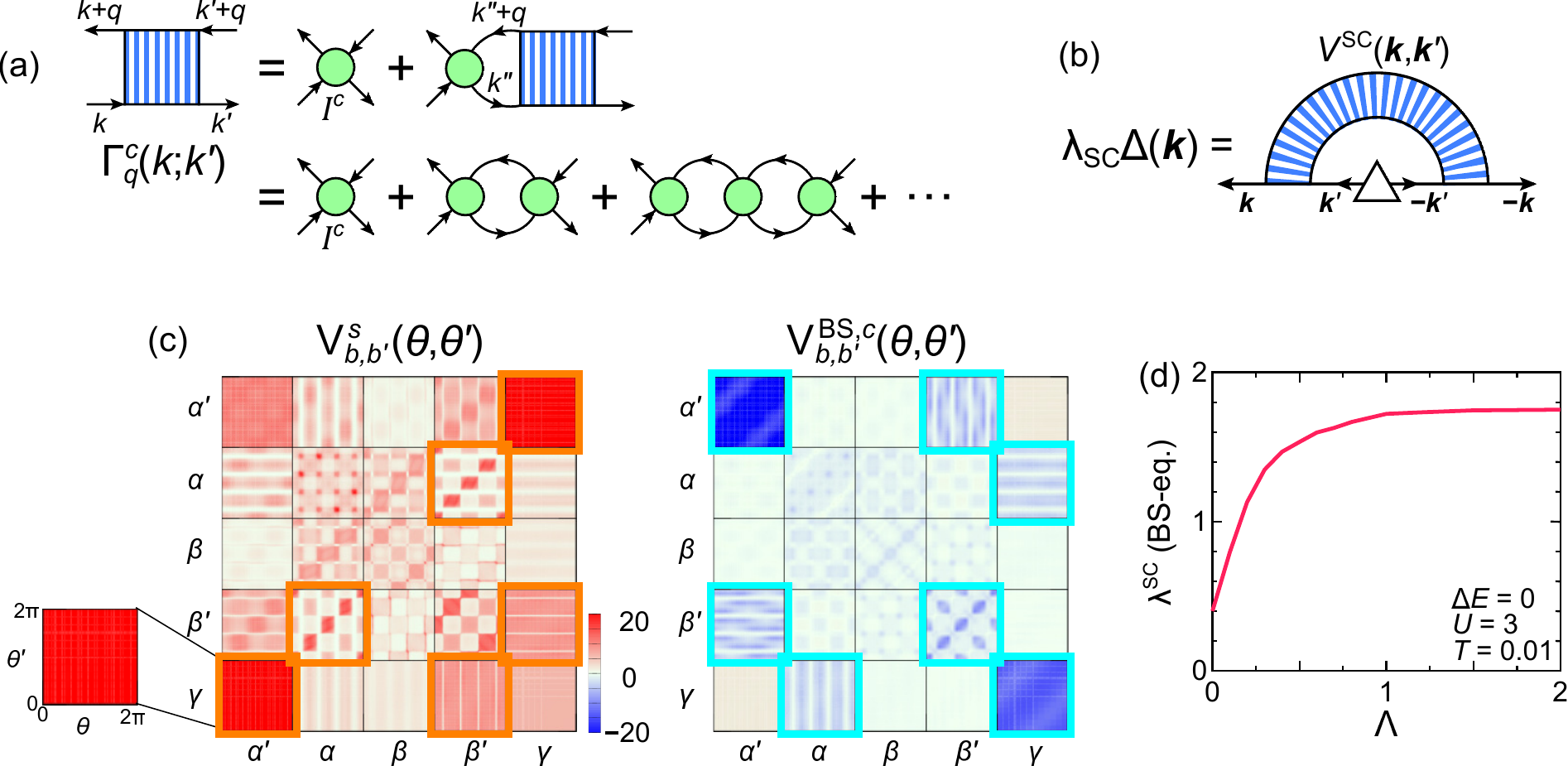}
\caption{
{\bf Superconducting pairing interaction in the BS-equation approach:} \ 
(a) Bethe--Salpeter equation for the full interaction vertex $\Gamma$.
(b) Linearized superconducting gap equation.
(c) Pairing interactions mediated by SDW and CDW fluctuations.
The orange and light blue solid boxes highlight the dominant repulsive interparity interactions mediated by SDW fluctuations and attractive intraparity interactions mediated by CDW fluctuations, respectively.
The strong repulsion between the opposite parity pockets arises from SDW fluctuations at $\q\sim{\bm Q}_{\rm sdw}, \ (\pi,\pi)$, whereas the strong attraction between the same parity pockets arises from CDW fluctuations at $\q\sim{\bm Q}_{\rm cdw}, \ (0,0)$; see Fig.~\ref{fig2} in the main text.
(d) Dependence of the superconducting eigenvalue on the energy cutoff $\Lambda$ used in the BS-equation calculation.
}
\label{fig10}
\end{figure*}

We next formulate the superconducting pairing interaction beyond the conventional Migdal approximation
\cite{T14-Yamakawa}.
The same charge-channel vertex corrections that strongly enhance the CDW fluctuations also generate a pronounced attractive pairing interaction.
In Sec.~\ref{sec:methodD}, we have introduced the linearized superconducting gap equation in the band basis as follows:
\begin{equation}
\begin{aligned}
&\lambda_{\rm SC}\bar{\Delta}_b({\bm k},\epsilon_n)
=-\frac{T}{N}\sum_{n'}\sum_{{\bm k}',b'}V^{\rm SC}_{b,b'}(k,k')
\\
&\qquad\times
\left|G_{b'}({\bm k}',\epsilon_{n'})\right|^2
\bar{\Delta}_{b'}({\bm k}',\epsilon_{n'}),
\end{aligned}
\label{eq:SI-gap}
\end{equation}
which is illustrated in Fig.~\ref{fig10}(b).
Here, $b$ and $b'$ are band indices, $k=({\bm k},\epsilon_n)$ and $k'=({\bm k}',\epsilon_{n'})$, and $G_b({\bm k},\epsilon_n)$ is the band Green function including the FLEX self-energy. The superconducting eigenvalue $\lambda_{\rm SC}$ reaches unity at $T=T_c$.
The quasiparticle gap function at zero energy is given by 
\begin{equation}
\Delta_b({\bm k})
\equiv
\frac{{\bar{\Delta}}_b^{R}({\bm k},0)}{Z_b({\bm k})}.
\label{eq:SI-physical-gap}
\end{equation}
where 
$\bar{\Delta}_b^{R}({\bm k},\epsilon)$ is the retarded function derived from the analytic continuation of $\bar{\Delta}_b({\bm k},i\epsilon_n)$.
$Z_b({\bm k})$ is the mass enhancement factor.

The singlet pairing interaction in the orbital--layer representation is written as
\begin{equation}
\hat V^{{\rm SC}}(k,k')= \hat V^{{\rm BS},c}(k,k') +\hat V^{s}(k,k')
+\frac{1}{2} (\hat C^s-\hat C^c).
\label{eq:SI-VSC}
\end{equation}
This interaction is transformed to the band representation, $V^{\rm SC}_{b,b'}(k,k')$, using the unitary matrix connecting the orbital--layer and band bases, and is then used in Eq.~(\ref{eq:SI-gap}).
Here, $\hat V^{{\rm BS},c}(k,k')$ represents the attractive interaction mediated by the charge channel fluctuations.
In the present system, $\hat V^{{\rm BS},c}(k,k')$ is strongly
enhanced by AL-type vertex corrections, which describe
quantum-interference effects.
In contrast, $\hat V^{s}(k,k')$ is well approximated by the FLEX formalism: $\frac{3}{2}\hat C^s\hat\chi^s(k-k')\hat C^s$.

In this paper, we evaluate $\hat V^{\rm BS,c}$ based on the Bethe--Salpeter (BS) equation method developed in Ref.~\cite{T14-Yamakawa}.
The charge-channel full four-point vertex satisfies
\begin{equation}
\begin{aligned}
\hat\Gamma_{{\bm q}}^{c}(k,k')
&=\hat I_{{\bm q}}^{c}(k,k')
-\frac{T}{N}\sum_p\hat I_{{\bm q}}^{c}(k,p)
\\
&\quad\times\hat G(p)\hat G(p+q)\hat\Gamma_{{\bm q}}^{c}(p,k').
\end{aligned}
\label{eq:SI-BS}
\end{equation}
When evaluating the BS vertex, we discretize the external momenta $\k$ and $\k'$ on the Fermi surfaces using 16 points per pocket.
Increasing the number of points per pocket to 32 changes the superconducting eigenvalue by only a few percent.
By contrast, the internal momentum--frequency sum in Eq.~(\ref{eq:SI-BS}) is evaluated with the momentum running over the entire Brillouin zone \cite{T14-Yamakawa}.
In our numerical implementation, we first compute $\tilde{\Gamma}_\q(k,p)$ with ${\bm p}$ and ${\bm p}+{\bm q}$ restricted to the Fermi surfaces, while ${\bm k}$ runs over the full momentum mesh.
We then obtain $\Gamma_\q(p',p)$ by evaluating $\tilde{\Gamma}_\q(k,p)$ at the Fermi momentum $\k={\bm p}'$ \cite{T14-Yamakawa}.


Then, the charge-channel pairing interaction entering Eq.~(\ref{eq:SI-VSC}) is obtained as
\begin{equation}
\hat V^{\rm BS, c}(k,k')
=
-\frac{1}{2}
\hat\Gamma_{{\bm k}'-{\bm k}}^{c}(k,-k')
+\frac{1}{2}\hat C^c,
\label{eq:SI-Vbond-BS}
\end{equation}
in which the first-order term in Coulomb interaction is cancelled.
The BS equation means that $V^{\rm BS,c}$ diverges when the corresponding DW eigenvalue reaches unity.
In our numerical calculations, we further include in $V^{\rm BS,c}$ the correction for double counting shown in Fig.~8 of Ref.~\cite{T14-Yamakawa}.
This correction corresponds to the cross term reported in Ref.~\cite{T27-Yamakawa} and provides a weak attractive interaction.

Near a CDW instability, $V^{\rm BS,c}(k,k')$ is well approximated by the leading bond-order mode. The $M_z$-parity selection rule for the pairing interaction discussed in this work follows directly from this leading-mode approximation:
\begin{equation}
V^{\rm BS,c}(k,k')
\approx
f_{{\bm k}-{\bm k}'}(k')
f_{{\bm k}'-{\bm k}}(-k')
\frac{g_{{\bm k}-{\bm k}'}}
{1-\lambda_{{\bm k}-{\bm k}'}},
\label{eq:SI-Vbond-ff}
\end{equation}
where $\lambda_{{\bm q}}$ and $f_{{\bm q}}(k)$ are the leading eigenvalue and form factor of the DW equation, respectively.
The coefficient $g_{{\bm q}}$ represents the projection of the irreducible kernel onto the corresponding DW form factor.
Since $|g_{{\bm q}}|\gtrsim U$ is generally satisfied \cite{T14-Yamakawa,T12-Tazai},
the pairing interaction is strongly enhanced as the CDW eigenvalue approaches unity.
Importantly, the full $V^{\rm BS,c}(k,k')$ used in our calculations includes contributions from all DW modes.

In the numerical implementation of the BS-equation method used in this work, we employ the following approximations for $V^{\rm BS,c}_{b,b'}(k,k')$.
(i) The vertex corrections are calculated only for the lowest fermionic Matsubara frequencies $\epsilon_n$ and $\epsilon_{n'}$.
For higher Matsubara frequencies, we neglect the vertex corrections and use the FLEX-level interaction.
(ii) The momentum dependence of $V^{\rm BS, c}_{b,b'}$ is calculated only on the Fermi surfaces and is extended to the states within the cutoff window according to
\begin{equation}
V^{\rm BS,c}_{b,b'}(k,k')
\approx
V^{\rm BS,c}_{b,b'}(\theta_{\bm k},\theta_{{\bm k}'})
\Theta_{b,{\bm k}}\Theta_{b',{\bm k}'}.
\label{eq:SI-Vbond-angle}
\end{equation}
Here, $\theta_{\bm k}=\tan^{-1}(\tilde{k}_y/\tilde{k}_x)$, where
$\tilde{\bm k}\equiv{\bm k}-{\bm k}_b$ and ${\bm k}_b$ denotes the center of the $b$th Fermi surface.
We take ${\bm k}_b=(0,0)$ for the $\alpha$ and $\alpha'$ Fermi surfaces, and
${\bm k}_b=(\pi,\pi)$ for the $\beta$, $\beta'$, and $\gamma$ Fermi surfaces.
(iii) We restrict this extension using the cutoff function
$\Theta_{b,{\bm k}}=\theta(\Lambda-|\epsilon_{b,{\bm k}}-\mu|)$,
where $\theta(x)$ is the Heaviside step function.
We use $\Lambda=0.3$ as a conservative cutoff: no band crossing occurs within this window, allowing the band index to be traced continuously.
As shown in Fig.~\ref{fig10}(d), increasing $\Lambda$ further enhances $\lambda_{\rm SC}$.
The chosen cutoff therefore deliberately underestimates the CDW-mediated pairing enhancement relative to larger cutoff windows in this implementation.
It demonstrates strong cooperative pairing with a restricted energy window; a more quantitative determination of $T_c$ requires further treatment of the off-Fermi-surface and frequency dependence of the vertex.

Figure~\ref{fig10}(c) shows the resulting pairing interaction for $U=3$ and $T=0.01$. SDW fluctuations generate strong repulsive interactions between Fermi pockets with opposite $M_z$ parities (orange solid boxes). These interactions are particularly strong between the $\alpha'$ and $\gamma$ pockets, and are also prominent between the $\alpha$ and $\beta'$ pockets and between the $\beta'$ and $\gamma$ pockets. 
The subleading spin-fluctuation mode also contributes repulsive
interactions between states with the same $M_z$ parity, including
scattering within the $\alpha'$ pocket. These contributions become
relatively weaker as $\alpha_S \to 1$.

By contrast, CDW fluctuations generate strong attractive interactions connecting states with the same $M_z$ parity (light blue solid boxes): within the $\alpha'$, $\gamma$ and $\beta'$ pockets, between the $\alpha'$ and $\beta'$ pockets, and between the $\alpha$ and $\gamma$ pockets. This complementary parity structure allows the dominant SDW- and CDW-mediated pairing interactions to cooperate in stabilizing the $s_{\pm}$-wave superconducting state.

\section{Numerical results without the $\gamma$ pocket ($\Delta E=-0.1$)}
\label{app:H}

\begin{figure*}[!tp]
\centering
\includegraphics[width=0.76\textwidth]{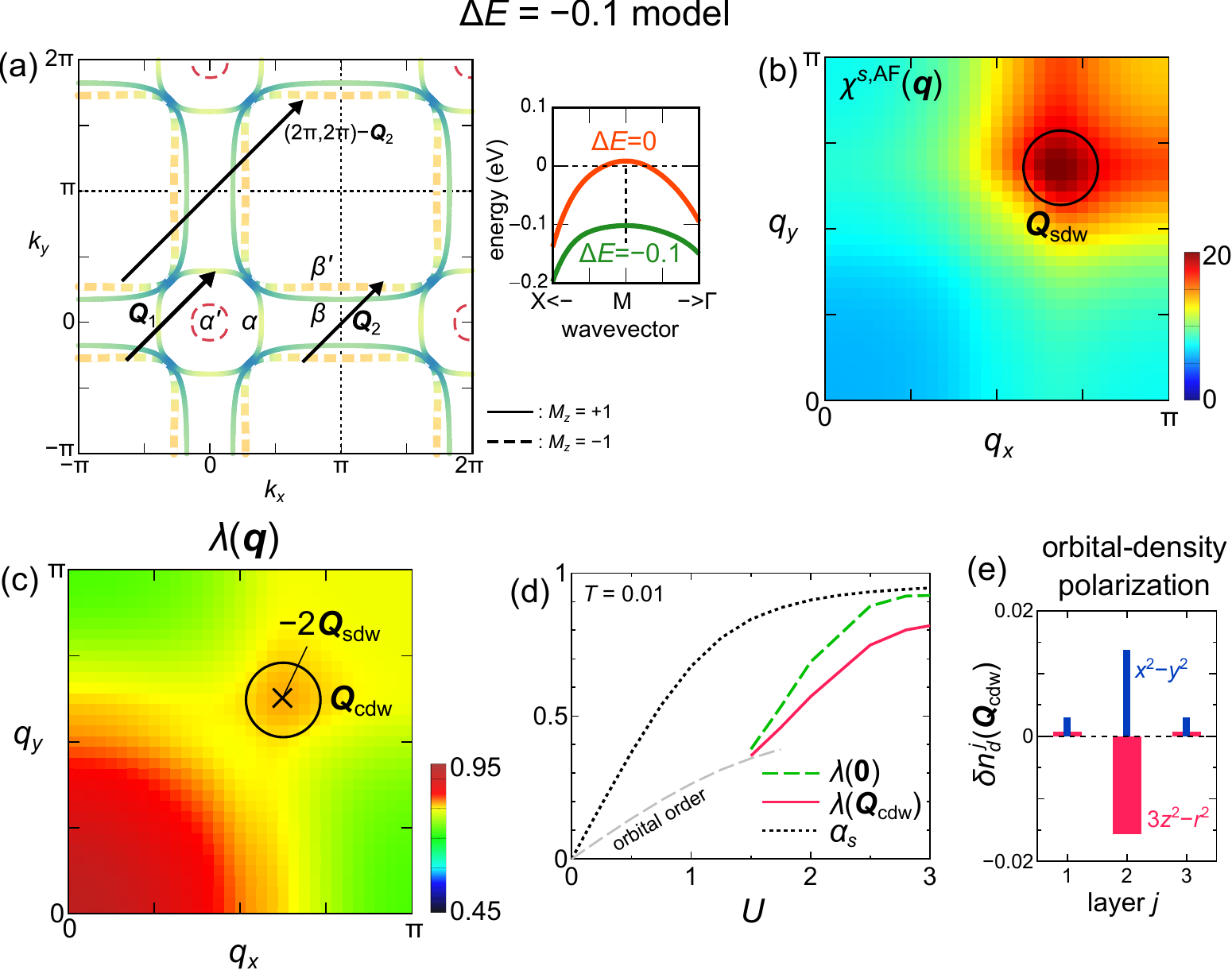}
\caption{
{\bf Robustness of the density-wave instability for $\Delta E=-0.1$:} \ 
(a) Fermi surfaces without the $\gamma$ pocket.
Solid and dashed lines denote the $M_z=+1$ and $-1$ sectors, respectively. 
(b) SDW susceptibility obtained by FLEX for $T=0.005$ and $U=3$.
(c) Charge-channel DW eigenvalue, showing robust structures near ${\bm Q}_{\rm cdw}$ and ${\bm q}={\bm 0}$.
(d) $U$-dependence of the charge-channel DW eigenvalue for $T=0.01$.
(e) Orbital-resolved density modulations induced by the leading CDW form factor for $\Delta E=-0.1$. 
}
\label{fig11}
\end{figure*}

We next examine the case $\Delta E=-0.1$, for which the top of the $\gamma$ band lies below the Fermi level and the $\gamma$ pocket disappears.
This calculation tests whether the quantum-interference-driven CDW fluctuations 
survive in the absence of the $\gamma$ pocket.

Figure~\ref{fig11}(a) shows the Fermi surfaces for $\Delta E=-0.1$, confirming the absence of the $\gamma$ pocket.
Figure~\ref{fig11}(b) shows the SDW susceptibility obtained by FLEX.
The peak position ${\bm Q}_{\rm sdw}$ is almost unchanged from those for $\Delta E=0$ and $0.1$ discussed in the main text.
Figure~\ref{fig11}(c) shows the CDW eigenvalue obtained from the DW equation.
It exhibits maxima near ${\bm Q}_{\rm cdw}$ and ${\bm q}={\bm 0}$, and the position of ${\bm Q}_{\rm cdw}$ is again almost unchanged from the cases $\Delta E=0$ and $0.1$.
Consequently, the characteristic relation
\begin{equation}
{\bm Q}_{\rm cdw}\approx2{\bm Q}_{\rm sdw}
\end{equation}
remains satisfied with good accuracy.
Figure~\ref{fig11}(d) exhibits the $U$-dependence of the charge-channel DW eigenvalue for $T=0.01$.
Thus, the emergence of the CDW instability due to the quantum-interference mechanism is robust against the disappearance of the $\gamma$ pocket.

\par\noindent\begin{minipage}{\columnwidth}
Figure~\ref{fig11}(e) shows the orbital-resolved density modulations for $\Delta E=-0.1$.
The response remains dominated by IL orbital polarization.
\end{minipage}\par

\FloatBarrier

\makeatletter
\let\auto@bib@innerbib\@empty
\makeatother

\end{document}